\documentclass[twocolumn,floatfix]{revtex4}
\pdfoutput=1
\usepackage{graphicx}
\usepackage{dcolumn}
\usepackage{longtable}
\usepackage{amsmath}
\usepackage{amssymb}

\begin{document}
\title{Parameter-free predictions for the collective deformation variables $\beta$ and $\gamma$ within the pseudo-SU(3) scheme
}

\author
{Dennis Bonatsos$^1$, Andriana Martinou$^1$, S. Sarantopoulou$^1$, I.E. Assimakis$^1$, S. K. Peroulis$^1$, and N. Minkov$^2$ }

\affiliation
{$^1$Institute of Nuclear and Particle Physics, National Centre for Scientific Research 
``Demokritos'', GR-15310 Aghia Paraskevi, Attiki, Greece}

\affiliation
{$^2$Institute of Nuclear Research and Nuclear Energy, Bulgarian Academy of Sciences, 72 Tzarigrad Road, 1784 Sofia, Bulgaria}

\begin{abstract}

The consequences of the short range nature of the nucleon-nucleon interaction, which forces the spatial part of the nuclear wave function to be as symmetric as possible, on the pseudo-SU(3) scheme are examined
through a study of the collective deformation parameters $\beta$ and  $\gamma$ in the rare earth region. It turns out that beyond the middle of each harmonic oscillator shell possessing an SU(3) subalgebra, the highest weight irreducible representation (the hw irrep) of SU(3) has to be used, instead of the irrep with the highest eigenvalue of the second order Casimir operator of SU(3) (the hC irrep), while in the first half of each shell the two choices are identical. The choice of the hw irrep predicts a transition from prolate to oblate shapes just below the upper end of the rare earth region, between the neutron numbers $N=114$ and 116 in the W, Os, and Pt series of isotopes, in agreement with available experimental 
information, while the choice of the hC irrep leads to  
a prolate to oblate transition in the middle of the shell, which is not seen experimentally. The prolate over oblate dominance in the ground states of even-even nuclei is obtained as a by-product. 

 \end{abstract}
 
\maketitle

\section{Introduction} 

Symmetries play an important role in shaping up the properties of atomic nuclei \cite{IA,IVI,FVI,Rosensteel,RW,Kota2}. On several occasions they impose specific forms of development of bands of energy levels as a function of the angular momentum, and/or strict selection rules on the electromagnetic transitions allowed among the energy levels in a given nucleus. Especially important are cases in which the predictions are parameter independent, since in these  comparison to the experimental data can lead to approval or rejection of a theory in a straightforward way. 

The symmetry of the nuclear wave function is made up by its spatial, spin, and isospin parts. Since the nucleons (protons and neutrons) are fermions, the total wave function has to be antisymmetric. At this point the short range nature \cite{Ring,Casten} of the nucleon-nucleon interaction plays an important role, namely it forces the spatial part of the wave function to be as symmetric as possible \cite{PVIWigner}. As a consequence, it also shapes up the spin-isospin part of the wave function, which has to correspond to the conjugate irreducible representation of the spatial part in order to guarantee the antisymmetric nature of the total wave function \cite{Wybourne,Kota}. 

These ideas have been recently tested in the framework of the proxy-SU(3) scheme \cite{proxy1,proxy2} for heavy deformed nuclei. Within this scheme the SU(3) symmetry of the harmonic oscillator \cite{Wybourne,Moshinsky}, exploited by Elliott \cite{Elliott1,Elliott2,Elliott3} for the description of sd shell nuclei, which is known to be broken by the strong spin-orbit interaction in higher shells \cite{MJ}, is recovered. In more detail, it is known \cite{MJ} that the spin-orbit interaction destroys the SU(3) symmetry of a given harmonic oscillator shell by forcing the Nilsson orbitals bearing the highest value of the total angular momentum $j$ to go into the shell below, while the orbitals bearing the highest value of the total angular momentum $j$ from the shell above invade the shell under consideration. The proxy-SU(3) scheme recovers approximately the SU(3)  
symmetry by replacing in each shell the intruder orbitals which have invaded the shell from above by the orbitals which have escaped from this shell into the lower one. The replacement is based on orbitals differing by $\Delta K [\Delta N \Delta n_z \Delta \Lambda]=0[110]$ in the standard Nilsson \cite{Nilsson1,Nilsson2} quantum numbers $N$ (the total number of quanta),  $n_z$
(the total number of quanta along the body-fixed $z$-axis), $\Lambda$ (the projection of the orbital angular momentum along the body-fixed $z$-axis), and $K$ (the projection of the total angular momentum along the body-fixed $z$-axis). These 0[110] pairs of orbitals are known to exhibit maximal spatial overlaps \cite{Karampagia}, similar to the ones of the Federman-Pittel pairs \cite{FP1,FP2,FP3}, which are known to play a crucial role in the development of nuclear deformation. Furthermore, they are known to be related to enhanced proton-neutron interaction, as indicated by double differences of experimental nuclear masses \cite{Cakirli1,Cakirli2}.

The short range of the nucleon-nucleon interaction leads to intriguing results within the proxy-SU(3) scheme, as discussed in detail in Ref. \cite{GC40}. It turns out that up to the middle of a given U(N) shell possessing an SU(3) subalgebra, the irreducible representation (irrep) containing the ground state band and in most cases additional bands, like the $\gamma_1$ band and the first $K=4$ band, corresponds both to the hC irrep, i.e., the irrep  with the highest eigenvalue of the second order Casimir operator $C_2^{SU(3}(\lambda,\mu)$, where $\lambda$ and $\mu$ are the Elliott quantum numbers \cite{Elliott1,Elliott2,Elliott3} characterizing the SU(3) irrep $(\lambda,\mu)$, and to the hw irrep, i.e., the irrep possessing the highest weight \cite{code}. Beyond the middle of the shell, however, this agreement is destroyed. The ground state band and its companions lie in the highest weight (hw) irrep of SU(3), which is not the same as the hC irrep carrying the maximum eigenvalue of $C_2^{SU(3}(\lambda,\mu)$, except in the case of the last four particles in the shell. 

This breaking of the particle-hole symmetry in the U(N) shells has two important physical implications: a) The majority of nuclei in a shell exhibit ground states with prolate deformations, thus resolving the long standing question \cite{Hamamoto} of the reason causing the dominance of prolate over oblate deformation in the ground states of even-even nuclei. b) A prolate to oblate transition is observed near the end of both the neutron 82-126 shell  and the proton 50-82 shell in rare earth nuclei, in agreement to experimental evidence \cite{Namenson,Alkhomashi,Wheldon,Podolyak,Linnemann} and theoretical predictions \cite{proxy2,proxy3}. 

The crucial role played by the short range of the interaction and the Pauli principle is also manifested in an another field of physics, namely metallic clusters \cite{deHeer,Brack,Nester,deHeer2}. The potentials used in metallic clusters have a harmonic oscillator-like shape similar to that of the nuclear potentials, albeit with a depth smaller by several orders of magnitude \cite{deHeer,Nester,deHeer2}. Metallic clusters are simpler, since in their case neither the spin-orbit interaction nor the pairing force are present \cite{Clemenger,Greiner}.  The short range nature of the interaction and the Pauli principle alone, within the SU(3) symmetry lead to experimentally observed  \cite{Borggreen,Pedersen1,Pedersen2,Haberland,Schmidt} prolate shapes above the magic numbers  \cite{Martin1,Martin2,Bjorn1,Bjorn2,Knight1,Peder,Brec1,Brec2}   seen in alkali clusters, while oblate shapes are seen below these magic numbers. Details of this study will be reported elsewhere \cite{PRA}. In the framework of this study \cite{PRA} it is also proved that the highest weight SU(3) irrep for a given number of particles is the irrep characterized by the highest percentage of symmetrized particles allowed by the Pauli principle in relation to the total number of particles in the physical system.  
Therefore we see that the short range of the interaction and the Pauli principle, when present in systems bearing an underlying SU(3) symmetry, have important consequences of global validity.     

In the present work we apply the particle-hole symmetry breaking to the pseudo-SU(3) scheme \cite{Adler,Shimizu,pseudo1,pseudo2,Ginocchio1,Ginocchio2}, in which the SU(3) symmetry of the harmonic oscillator \cite{Wybourne,Moshinsky} is restored by replacing the quantum numbers (number of quanta, angular momenta, spin) characterizing the levels remaining in a nuclear shell after the desertion of the levels going into the shell below by new, ``pseudo'' quantum numbers, which map these levels onto a full shell with one quantum of excitation less. In this way the ``remaining'' levels (also called normal parity levels) acquire full SU(3) symmetry, while the intruder levels (also called abnormal parity levels, since they have parity opposite to the parity of the ``remaining'' ones) are set aside and treated separately. It should be noted that the replacement of the ``remaining'' levels by their ``pseudo'' counterparts is an exact process, described by a unitary transformation \cite{Quesne,Velazquez}. In the lowest order approximation, the intruder levels are treated as spectators, while the ``remaining'' levels, as described in the framework of the pseudo-SU(3) symmetry, derive the collective properties of the nucleus. This is the approximation we are going to use in the present work. At a more sophisticated level, the intruder levels are taken into account \cite{DW1,DW2} as a single $j$-shell of the shell model.   

\section{SU(3) irreps in the pseudo-SU(3) scheme}

The first step needed for the description of a nucleus within the pseudo-SU(3) scheme is the distribution of the valence protons (neutrons) into the appropriate pseudo-SU(3) shells and the intruder orbitals. One way to achieve this is by looking at the relevant Nilsson diagrams \cite{Nilsson1,Nilsson2,Lederer}. If one knows the deformation of the nucleus, it is easy to place the valence protons (neutrons) in the  Nilsson orbitals of the relevant shell and see how many of them belong to the ``remaining'' orbitals and how many go into the ``intruder'' ones. In order to do so, one should have in hand in advance an estimate for the deformation of the given nucleus. We are going to use as such estimates the deformations predicted by the D1S Gogny interaction \cite{Gogny}, which are known to be in good agreement with existing experimental data \cite{Pritychenko}. The distribution of protons and neutrons into pseudo-SU(3) shells and intruder orbitals for the rare earth nuclei with $Z=50-82$ and $N=82-126$ is shown in Table 1. 

Once the number of valence protons (neutrons) in the relevant pseudo-SU(3) shell is known, the SU(3) irrep 
$(\lambda_p,\mu_p)$ characterizing the protons and the SU(3) irrep $(\lambda_n,\mu_n)$ characterizing the neutrons can be found by looking at Table 2, in which the SU(3) irreps corresponding to the relevant particle number are given. The valence protons of this shell live within a U(10) pseudo-SU(3) shell, which can accommodate a maximum of 20 particles, while the valence neutrons live within a U(15) pseudo-SU(3) shell, which can accommodate a maximum of 30 particles. In Table 2 for each U(N) algebra and each particle number, two irreps are reported. One of them is the highest weight irrep, labelled by hw, while the other is the irrep possessing the highest eigenvalue of the second order Casimir operator $C_2^{SU(3}(\lambda,\mu)$, labelled by hC. Both of them have been obtained using the code of Ref. \cite{code}. Details on the derivation and the physical meaning of the hw and hC irreps have been given in \cite{HNPS2017,Rila2018}.

As we have already remarked, we are going to produce results for two cases: a) both protons and neutrons live in shells in which the hw irreps are taken into account, b) both protons and neutrons live in shells in which the hC irreps are taken into account. In both cases it will be assumed that the whole nucleus is described by the ``stretched'' SU(3) irrep   $(\lambda_p+\lambda_n,\mu_p+\mu_n)$ \cite{DW1}. The results for the former case are shown in Table 3, while the results for the latter case are shown in Table 4. 

The procedure described above can be clarified by two examples. 

a) For $^{158}_{64}$Gd$_{94}$ one sees in \cite{Gogny} that the expected deformation is $\beta=0.356$, 
which is very close to the experimental value of 0.351 reported in \cite{Pritychenko}. 
The deformation parameter $\epsilon$ of the Nilsson model is related to $\beta$ through the equation 
$\epsilon=0.946 \beta$ \cite{Nilsson2}. Looking at the standard proton Nilsson diagrams \cite{Lederer} for $\epsilon=0.33$ we see that the 14 valence protons of $^{158}$Gd are occupying 5 orbitals of normal parity 
and 2 orbitals of abnormal parity. Similarly, looking at the standard neutron Nilsson diagrams \cite{Lederer} for $\epsilon=0.33$ we see that the 12 valence neutrons of $^{158}$Gd are occupying 4 orbitals of normal parity 
and 2 orbitals of abnormal parity. These values are reported in Table 1, taking into account that each orbital accommodates two particles. Now from Table 2 one sees that 
10 protons (the ones with normal parity) in the U(10) shell correspond to the irrep (10,4), while 
8 neutrons (the ones with normal parity) in the U(15) shell correspond to the irrep (18,4). Therefore the total irrep for $^{158}$Gd, reported in Table 3, is (28,8). Notice that since both the valence protons and neutrons of normal parity lie within the first half of their corresponding shell, the choice of the hw irreps vs. the hC irreps makes no difference, thus the same total irrep appears also in Table 4.   

b)For $^{176}_{70}$Yb$_{106}$ one sees in \cite{Gogny} that the expected deformation is $\beta=0.335$, 
which is close to the experimental value of 0.3014 reported in \cite{Pritychenko}. 
 Looking at the standard proton Nilsson diagrams \cite{Lederer} for $\epsilon=0.30$
 we see that the 20 valence protons of $^{176}$Yb are occupying 6 orbitals of normal parity 
and 4 orbitals of abnormal parity. Similarly, looking at the standard neutron Nilsson diagrams \cite{Lederer} for $\epsilon=0.30$ we see that the 24 valence neutrons of $^{176}$Yb are occupying 8 orbitals of normal parity 
and 4 orbitals of abnormal parity. These values are reported in Table 1. Now from Table 2 one sees that 
12 protons (the ones with normal parity) in the U(10) shell correspond to the hw irrep (12,0),
but they belong to the (4,10) hC irrep if the highest eigenvalue of $C_2^{SU(3)}$ is considered. Furthermore
16 neutrons (the ones with normal parity) in the U(15) shell correspond to the hw irrep (18,8),
but they belong to the (6,20) hC irrep if the highest eigenvalue of $C_2^{SU(3)}$ is considered.
Therefore the total irrep for $^{176}$Yb is (30,8) if the hw irreps are taken into account and is reported in Table 3. However, if the hC irreps are considered,
the total irrep for the same nucleus is (10,30), reported in Table 4. Notice that since both the valence protons and neutrons of normal parity lie within the second half of their corresponding shell, the choice of the hw irreps vs. the irreps with highest eigenvalue of $C_2^{SU(3)}$ makes a big difference, resulting in a clearly prolate $(\lambda > \mu)$ irrep in the first case and in a clearly oblate $(\lambda < \mu)$ irrep in the second case.  

\section{Shape variables} 

The labels $\lambda$ and $\mu$ of the SU(3) irrep $(\lambda,\mu)$ are known \cite{Castanos,Park} to be related to the shape variables $\beta$ and $\gamma$ of the geometric collective model of Bohr and Mottelson
\cite{BM}, in which $\beta$ describes the degree of deviation from the spherical shape (which corresponds to $\beta=0$), while $\gamma$ describes the deviation from axial symmetry, with $\gamma=0$ corresponding to prolate (rugby ball like) shapes, $\gamma=\pi/3$ corresponding to oblate (pancake like) shapes, and in between values corresponding to triaxial shapes, with maximum triaxality obtained at $\gamma=\pi/6$. The correspondence is obtained by mapping the invariants of the two theories onto each other \cite{Castanos,Park}. In more detail, the second and third order Casimir operators of SU(3), which are the invariants of the SU(3) model, are mapped onto the invariants of the collective model, which are $\beta^2$ and $\beta^3 \cos 3\gamma$. The results of the mapping are 
 \cite{Castanos,Park}
\begin{equation}\label{g1}
\gamma = \arctan \left( {\sqrt{3} (\mu+1) \over 2\lambda+\mu+3}  \right),
\end{equation}
and \cite{Castanos,Park}
\begin{equation}\label{b1}
	\beta^2= {4\pi \over 5} {1\over (A \bar{r^2})^2} (\lambda^2+\lambda \mu + \mu^2+ 3\lambda +3 \mu +3), 
\end{equation}
where the quantity in parentheses is related to the second order Casimir operator of SU(3) \cite{IA}, 
 \begin{equation}\label{C2} 
 C_2(\lambda,\mu)= {2 \over 3} (\lambda^2+\lambda \mu + \mu^2+ 3\lambda +3 \mu), 
\end{equation}
while $A$ is the mass number of the nucleus and $\bar{r^2}$ is related to the dimensionless mean square radius \cite{Ring}, $\sqrt{\bar{r^2}}= r_0 A^{1/6}$.
The constant $r_0$ is determined from a fit over a wide range of nuclei \cite{DeVries,Stone}. We use the value in Ref. \cite{Castanos}, $r_0=0.87$, in agreement to Ref. \cite{Stone}. 
 
The question of rescaling $\beta$ in relation to the size of the shell appears at this point, as in the case of proxy-SU(3) \cite{proxy2}. On one hand, from Table 2 (as well as from more extended tables reported in Ref. \cite{proxy2}), it is clear that $\lambda$ is proportional to the size of the shell, especially for irreps with $\lambda \gg \mu$. On the other hand, from Eq. (\ref{b1}) it becomes evident that for the same kind of irreps, $\beta$ is proportional to $\lambda$. If we had a theory in which all nucleons could be accommodated within a single SU(3) irrep, $\lambda$ would have been proportional to $A$. Since this is not the case here, it turns out that $\beta$ has to be multiplied by a factor $A/(S_p+S_n)$, where $S_p$ ($S_n$) is the size of the SU(3) proton (neutron) shell. In the present case, within the pseudo-SU(3) model, these sizes are 20 and 30 respectively, thus $\beta$ has to be multiplied by $A/50$. 

It would be instructive to compare the theoretical predictions to empirical information, where available. Experimental values for $\beta$ are determined by measuring the electric quadrupole transition rate from the $0_1^+$ ground state to the first excited $2_1^+$ state, $B(E2; 0_1^+\to 2_1^+)$, and are readily available in Ref. \cite{Pritychenko}. Since no direct experimental information exists for $\gamma$, we proceed, as in Ref. \cite{proxy2}, to an estimation of it from the energy ratio of the bandhead of the $\gamma_1$ band over the first excited state of the ground state band, 
\begin{equation}
R={E(2^+_2)\over E(2^+_1)}, 
\end{equation}
using as a bridge the Davydov model for triaxial nuclei \cite{DF,Casten,Esser}, in the framework of  which one has 
\begin{equation}\label{gm}
\sin 3\gamma= {3\over 2\sqrt{2}} \sqrt{1-\left({R-1\over R+1}  \right)^2}. 
\end{equation}
Empirical values of $\gamma$ extracted in this way will be used in the next section.  
 
\section{Numerical results}\label{num}

\begin{figure*}[htb]

{\includegraphics[width=60mm]{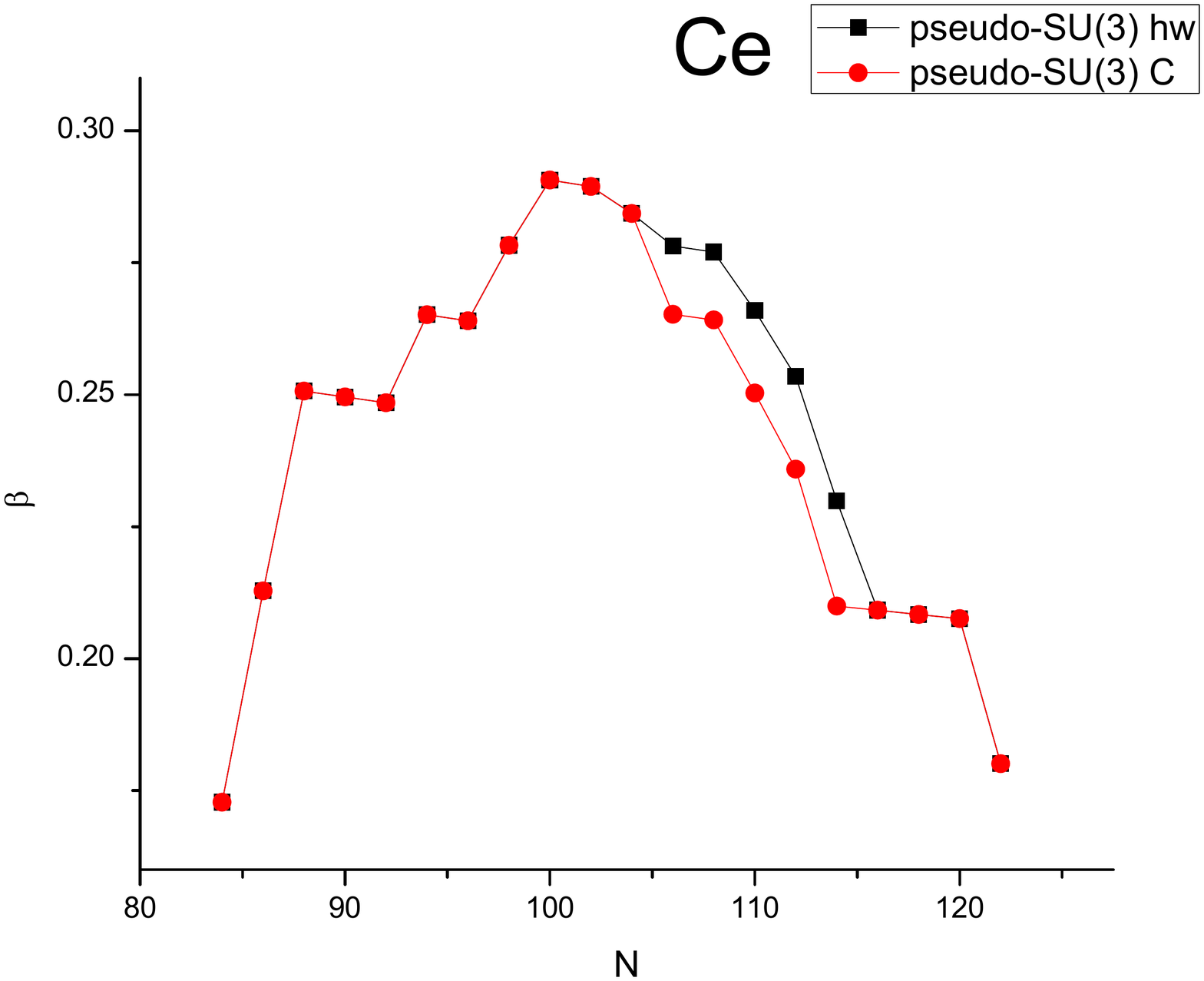}
\includegraphics[width=60mm]{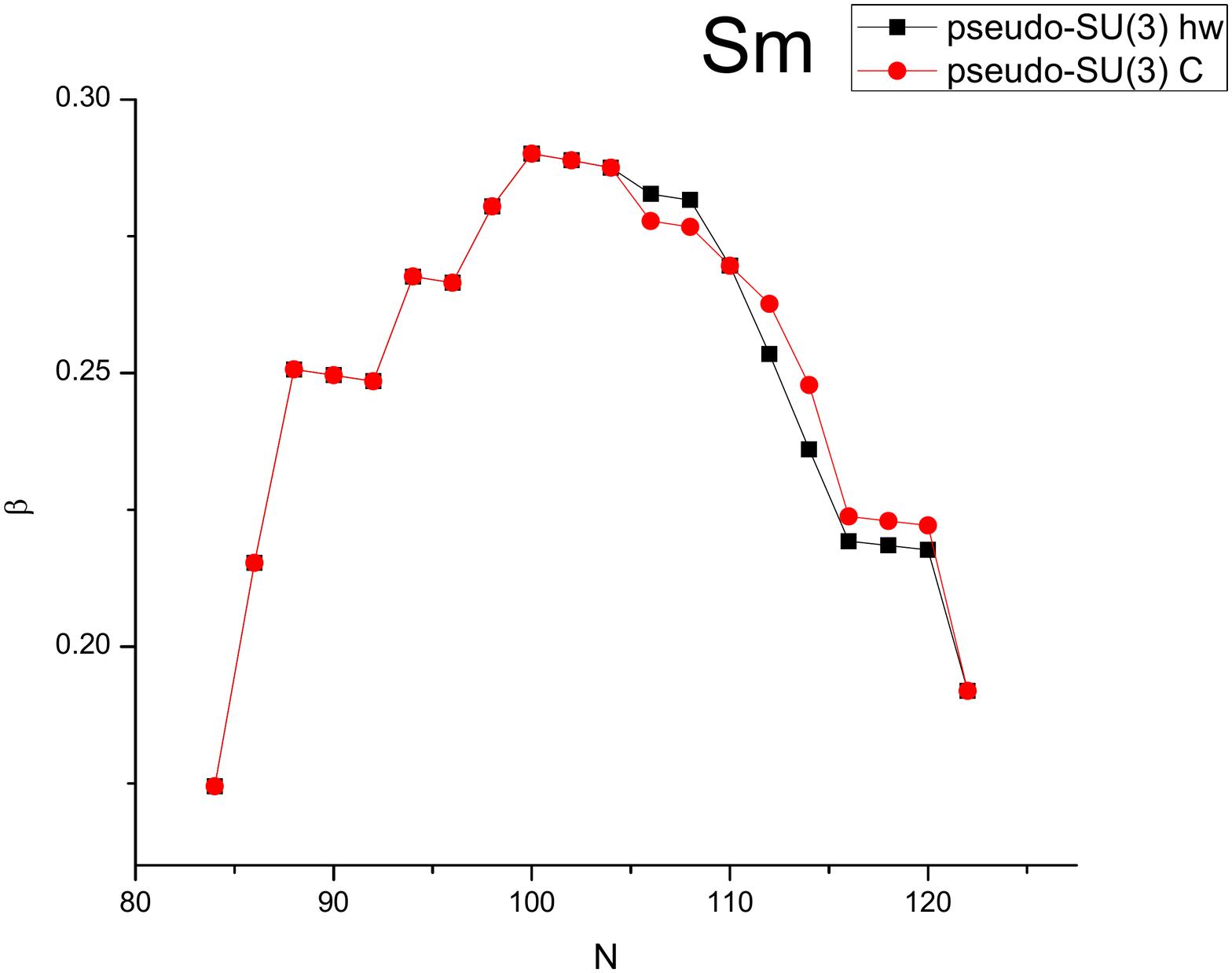}}
{\includegraphics[width=60mm]{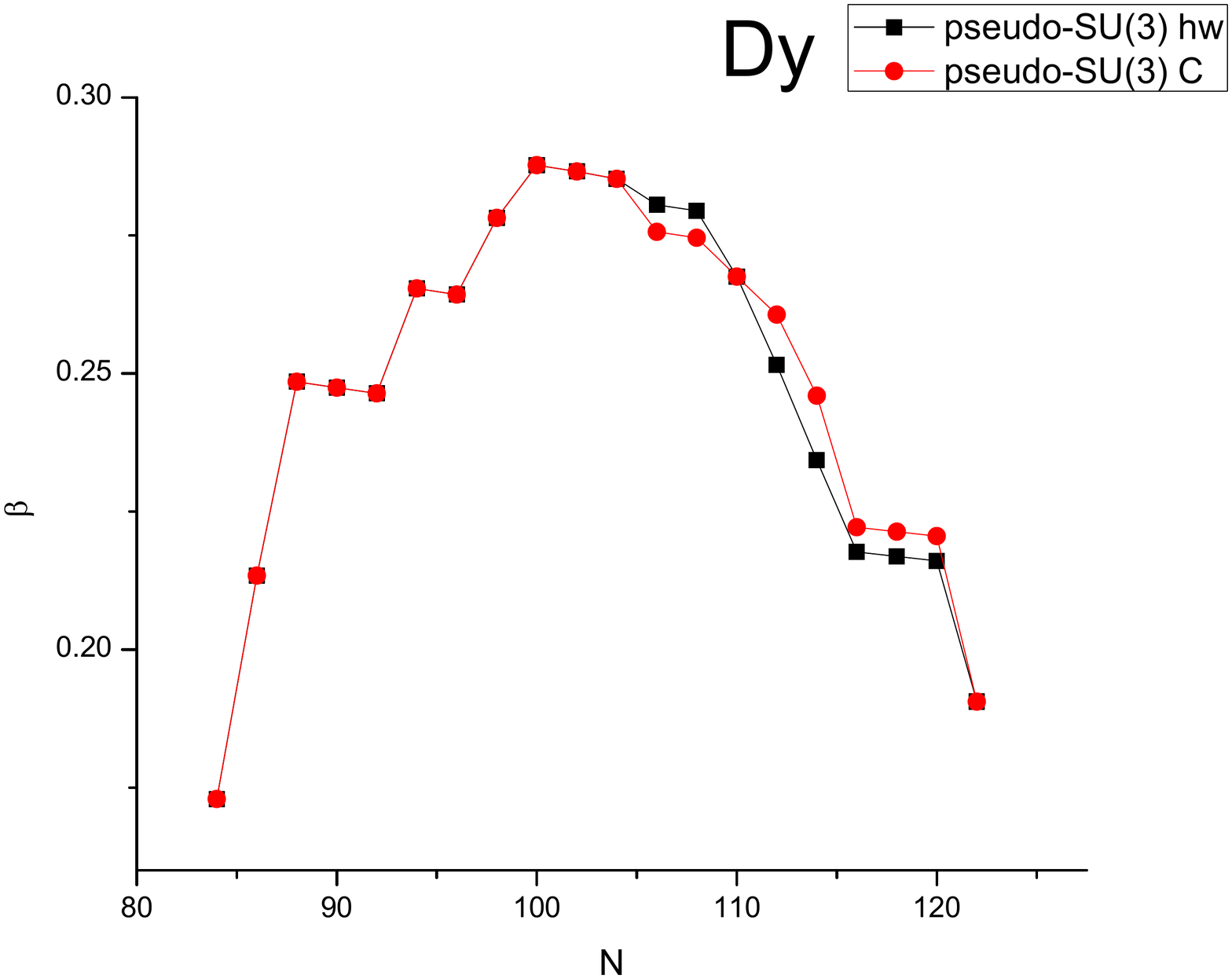}
\includegraphics[width=60mm]{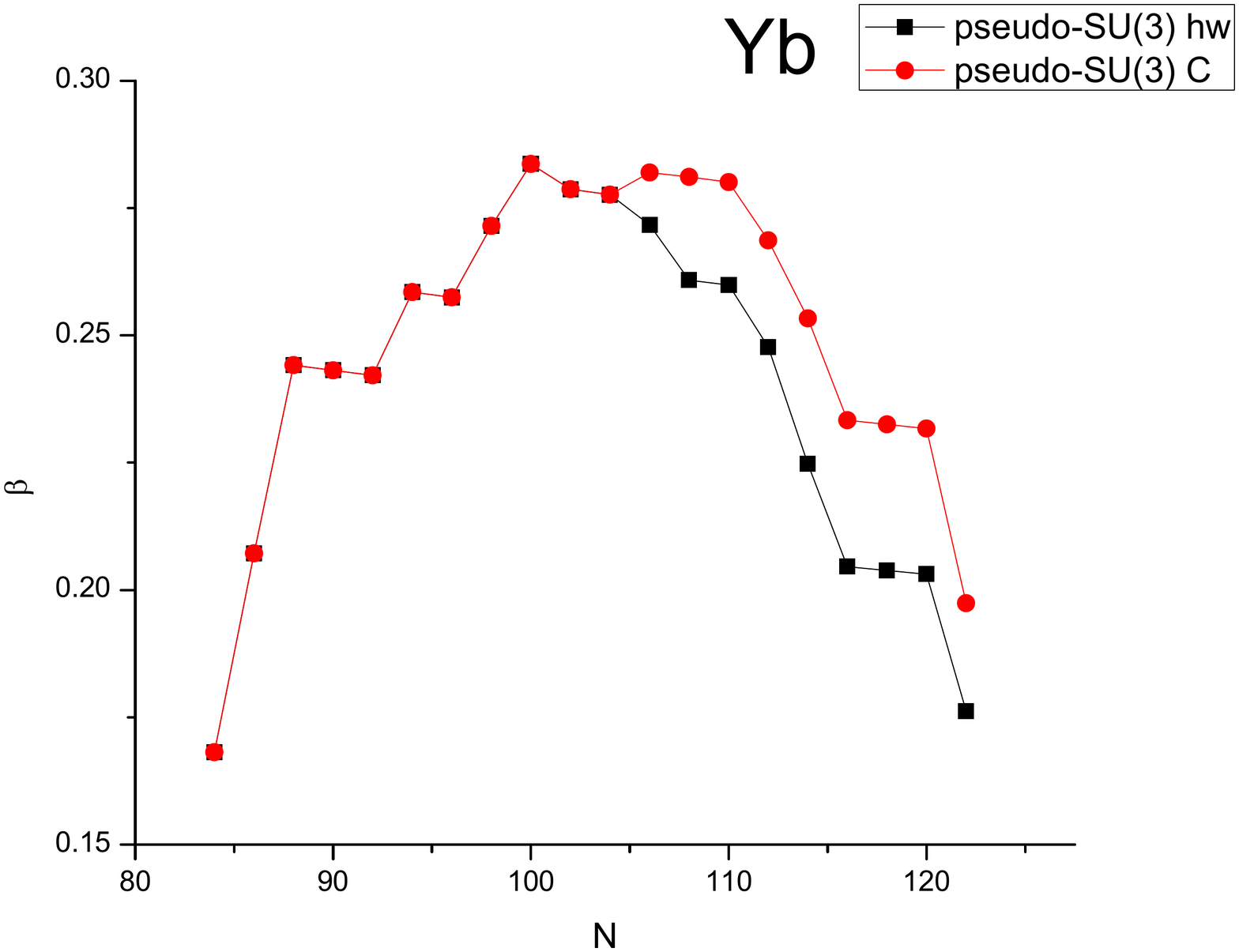}}
{\includegraphics[width=60mm]{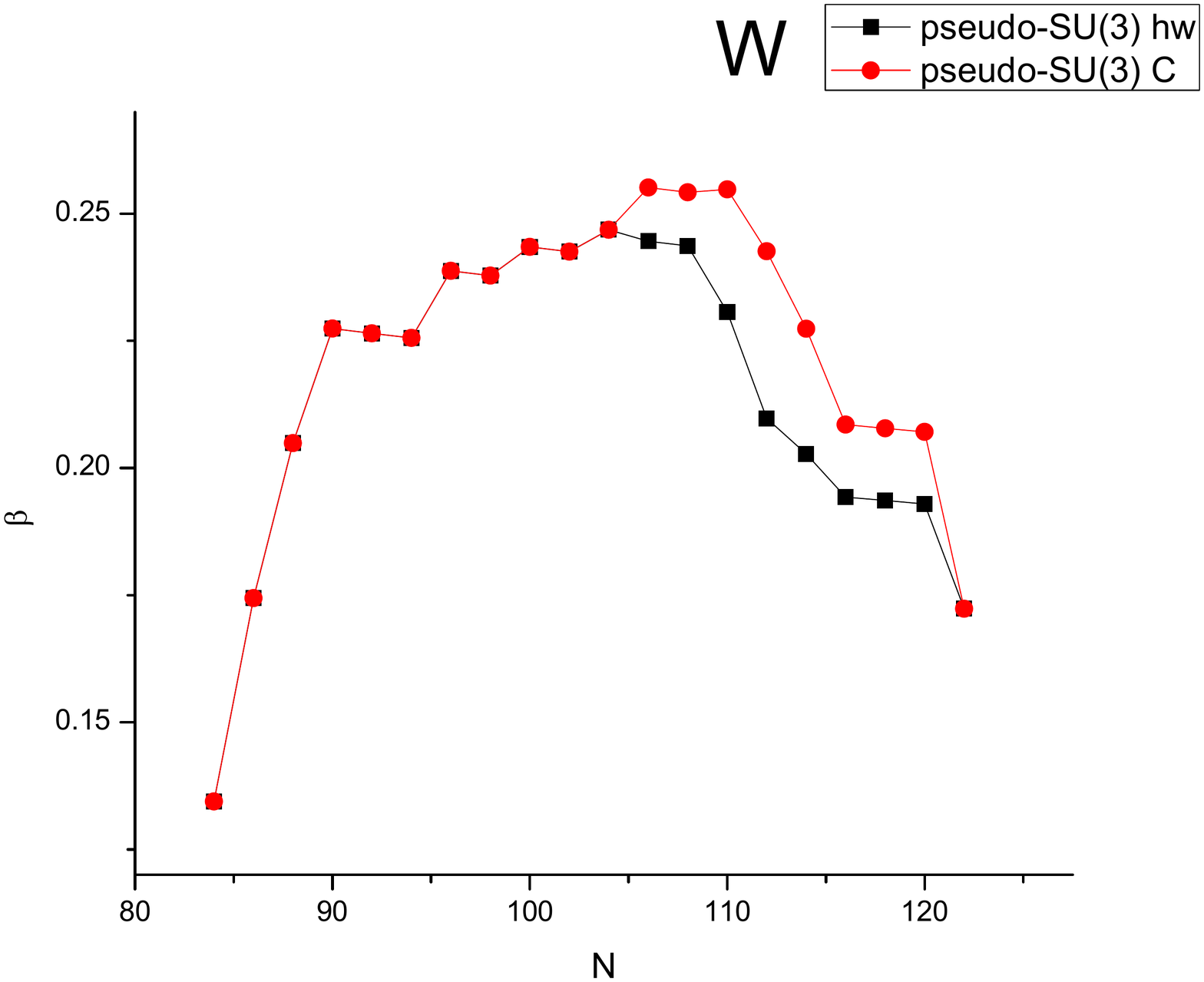}
\includegraphics[width=60mm]{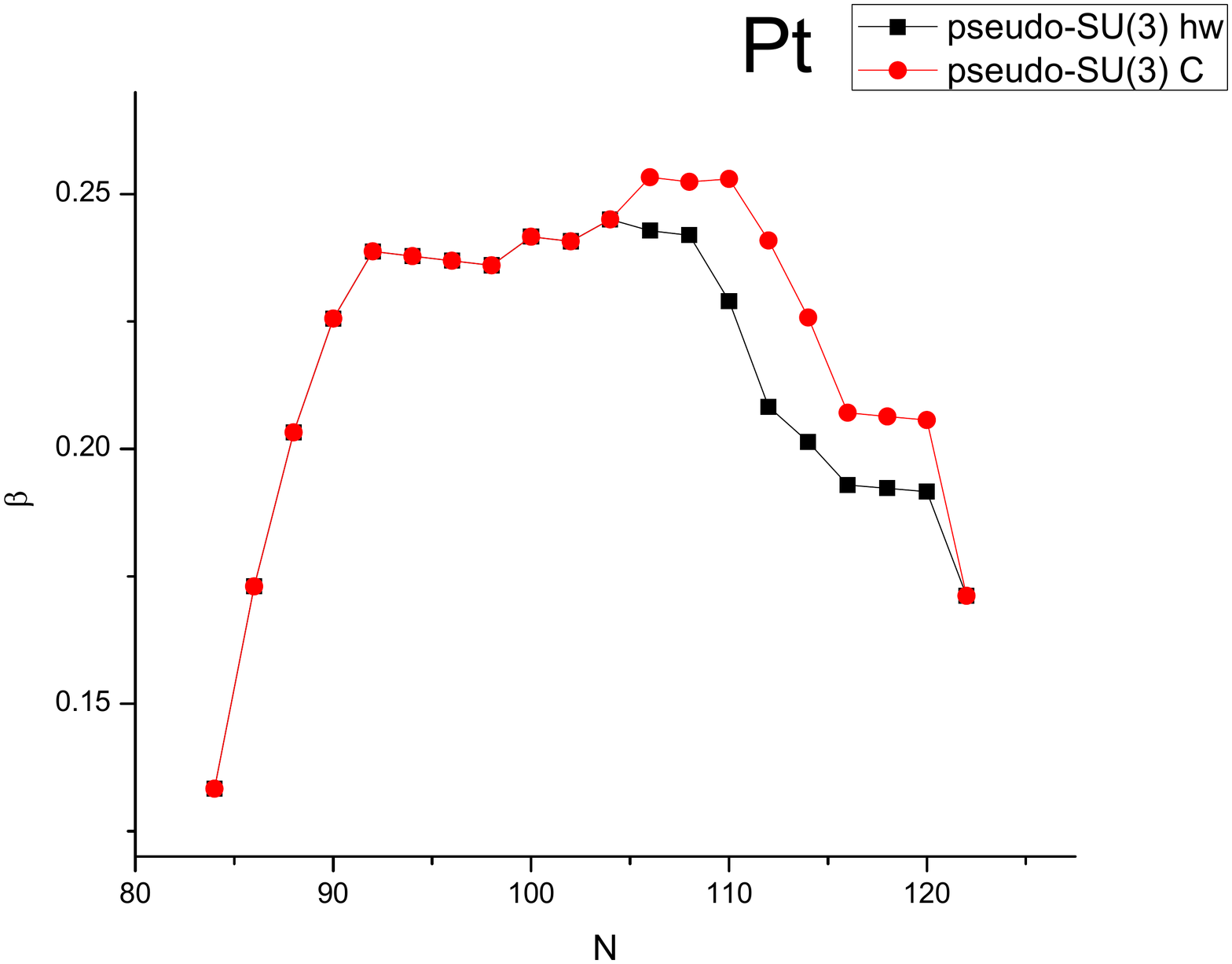}}

\caption{Pseudo-SU(3) predictions for the collective deformation variable $\beta$ for six series of isotopes in the rare earth region. The predictions labeled by hw have been obtained using the highest weight irreps of SU(3), 
while those labeled by C have been obtained using the hC irreps of SU(3) having the highest eigenvalue 
of the second order Casimir operator of SU(3), $C_2^{SU(3)}$. 
 See Section \ref{num} for further discussion.} 
\label{2B}
\end{figure*}

\begin{figure*}[htb]

\includegraphics[width=120mm]{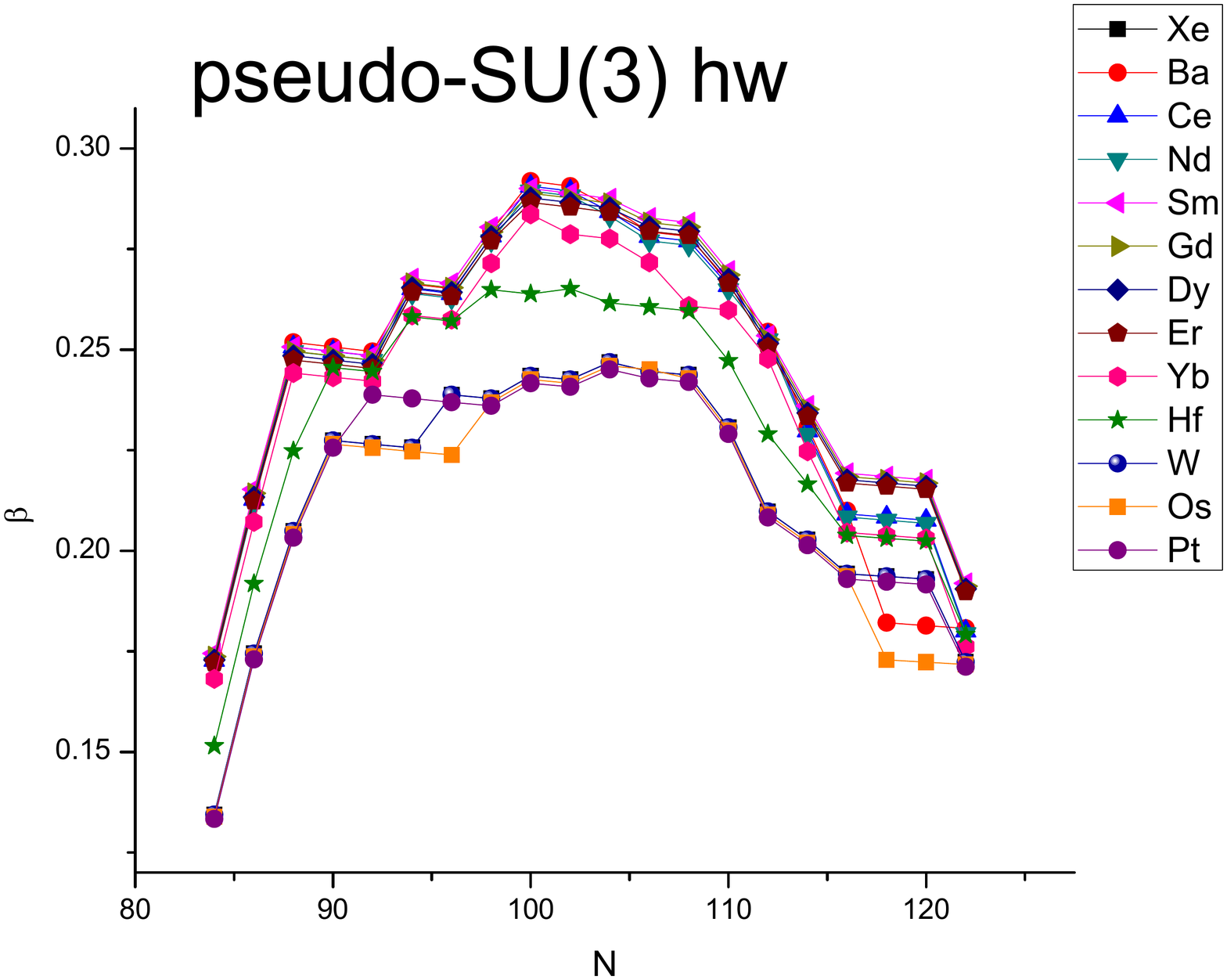} 
\includegraphics[width=120mm]{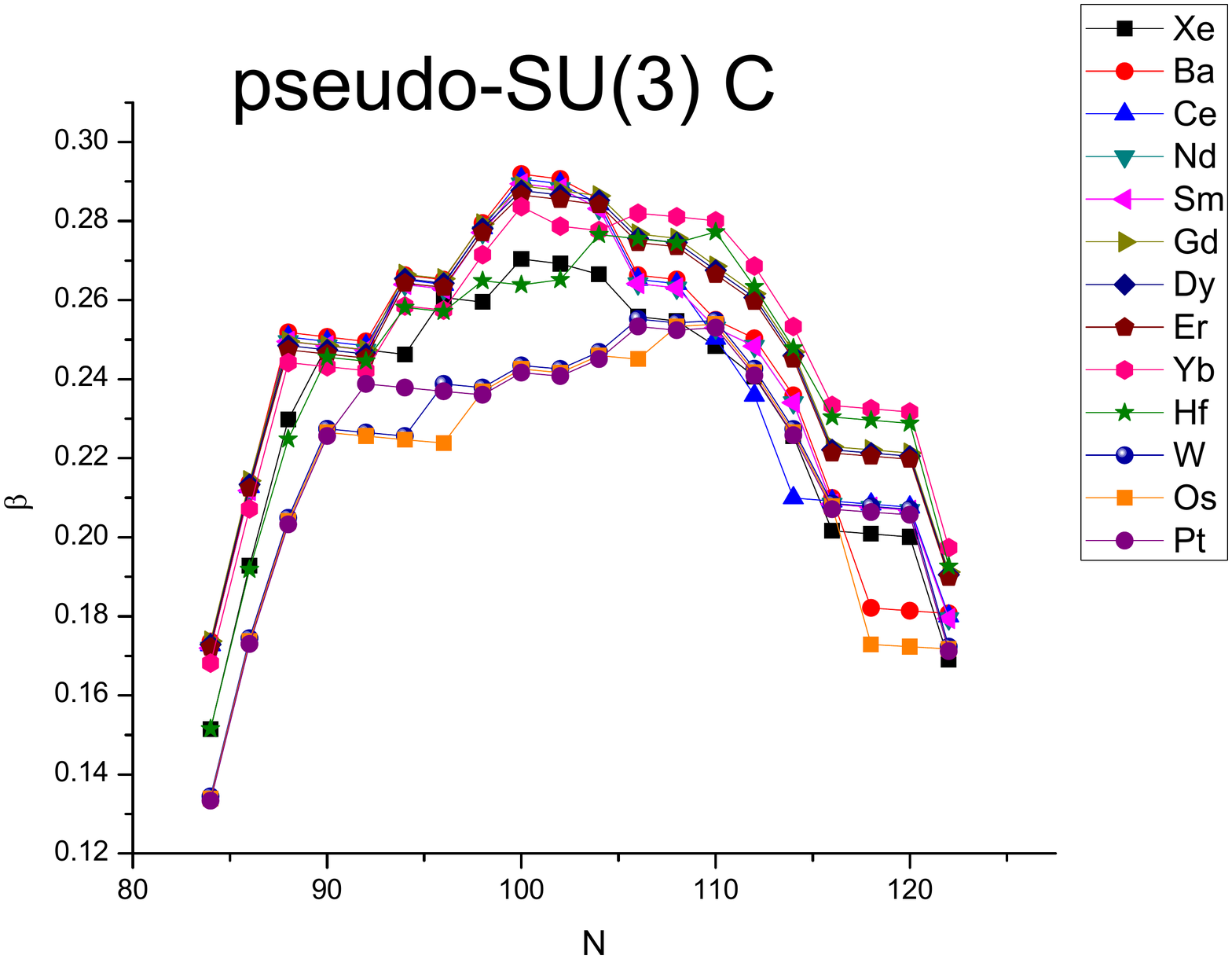}

\caption{Pseudo-SU(3) predictions for the collective deformation variable $\beta$ for the Xe-Pt series of isotopes in the rare earth region. The predictions labeled by hw (top panel) have been obtained using the highest weight irreps of SU(3), 
while those labeled by C (bottom panel) have been obtained using the hC irreps of SU(3) having the highest eigenvalue 
of the second order Casimir operator of SU(3), $C_2^{SU(3)}$. 
 See Section \ref{num} for further discussion.} 

\label{B}
\end{figure*}

\begin{figure*}[htb]

{\includegraphics[width=60mm]{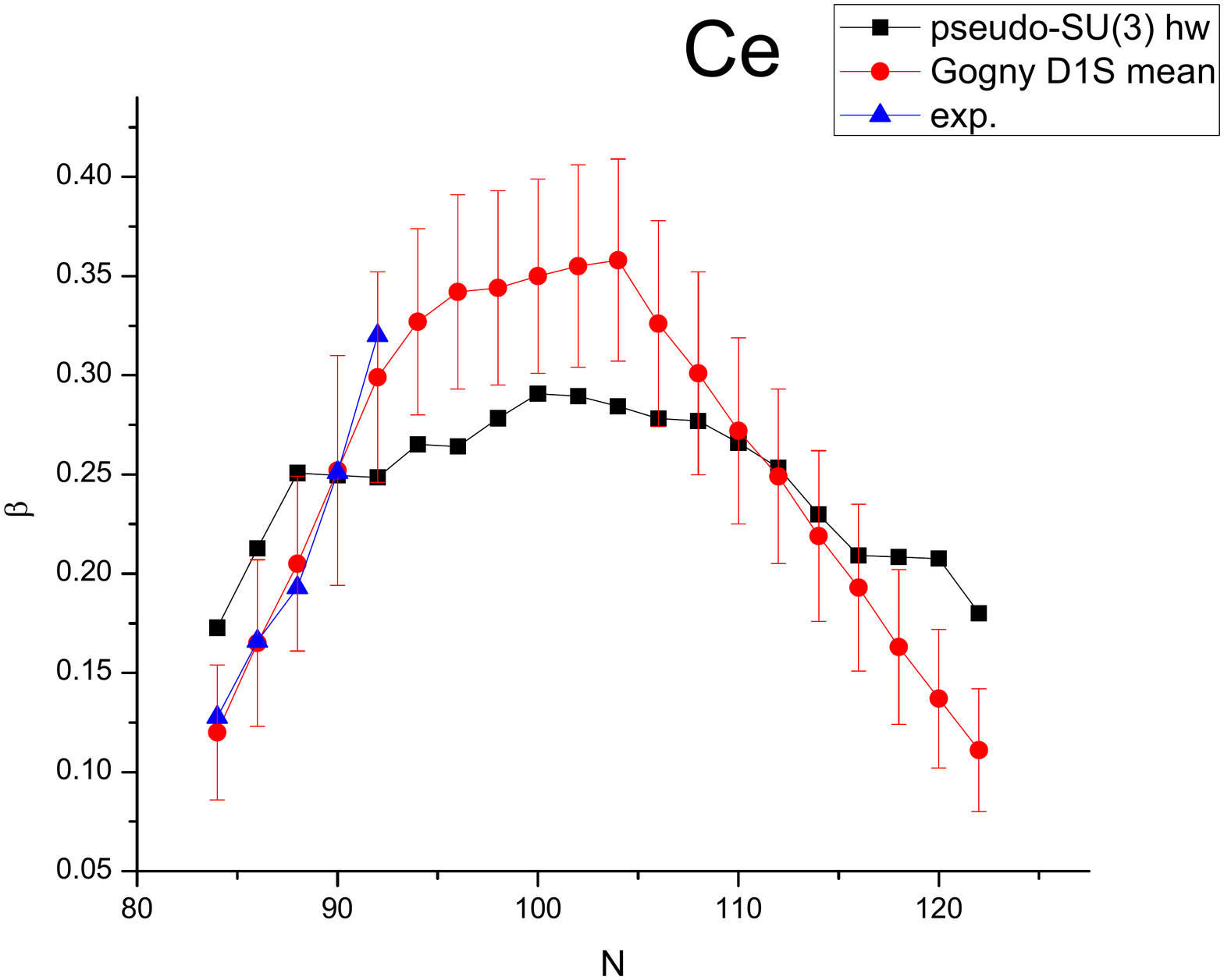}
\includegraphics[width=60mm]{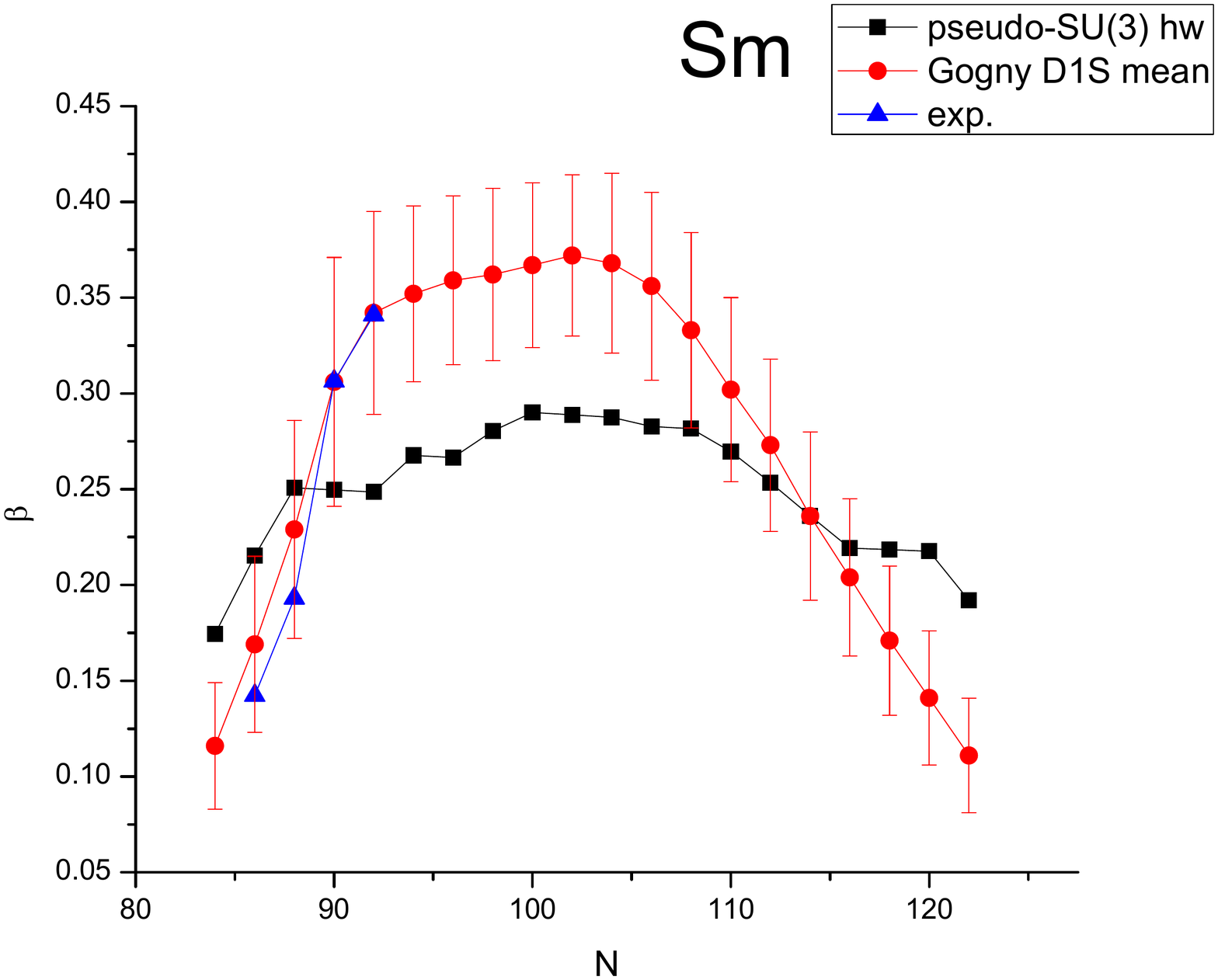}
\includegraphics[width=60mm]{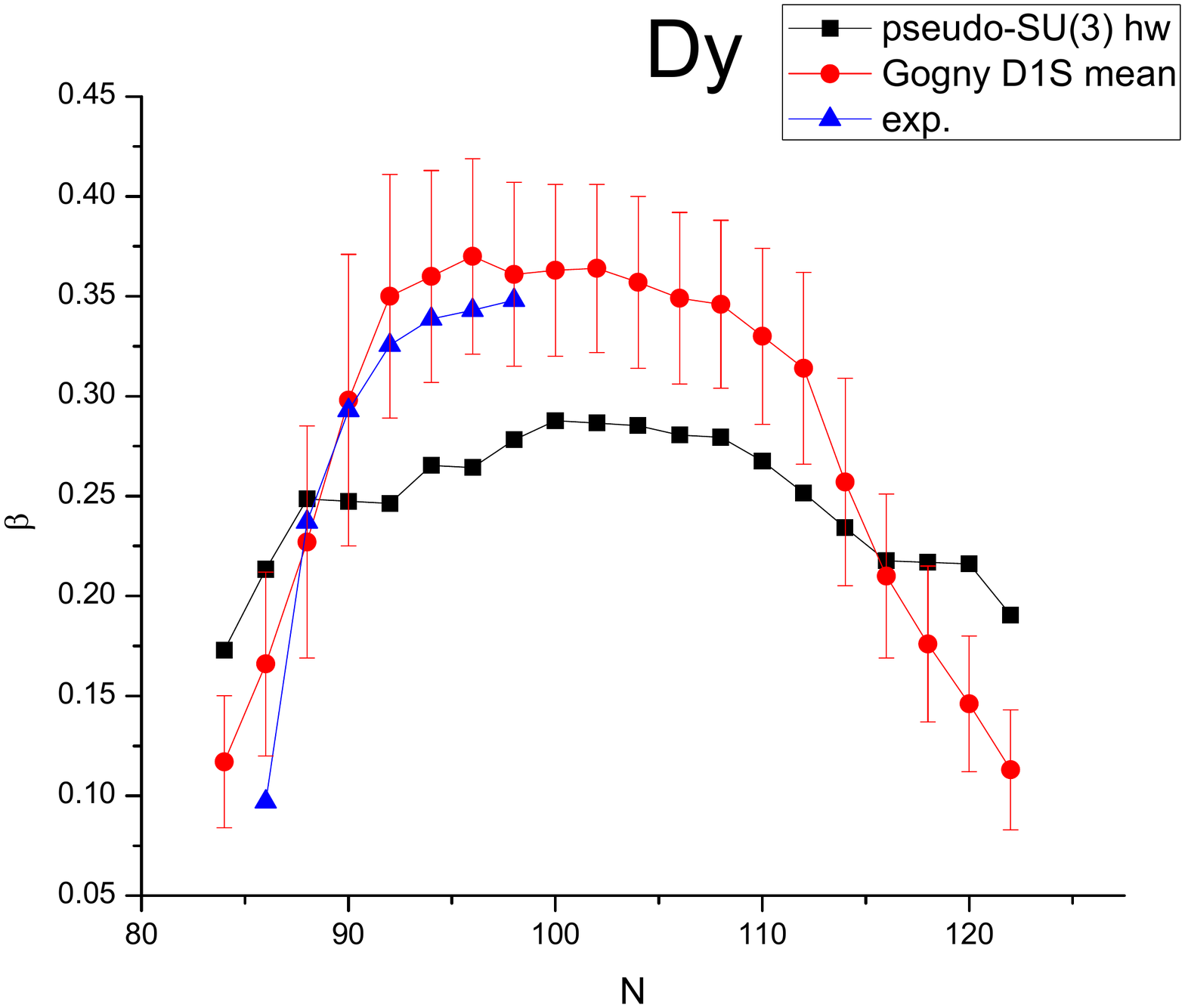}}
{\includegraphics[width=60mm]{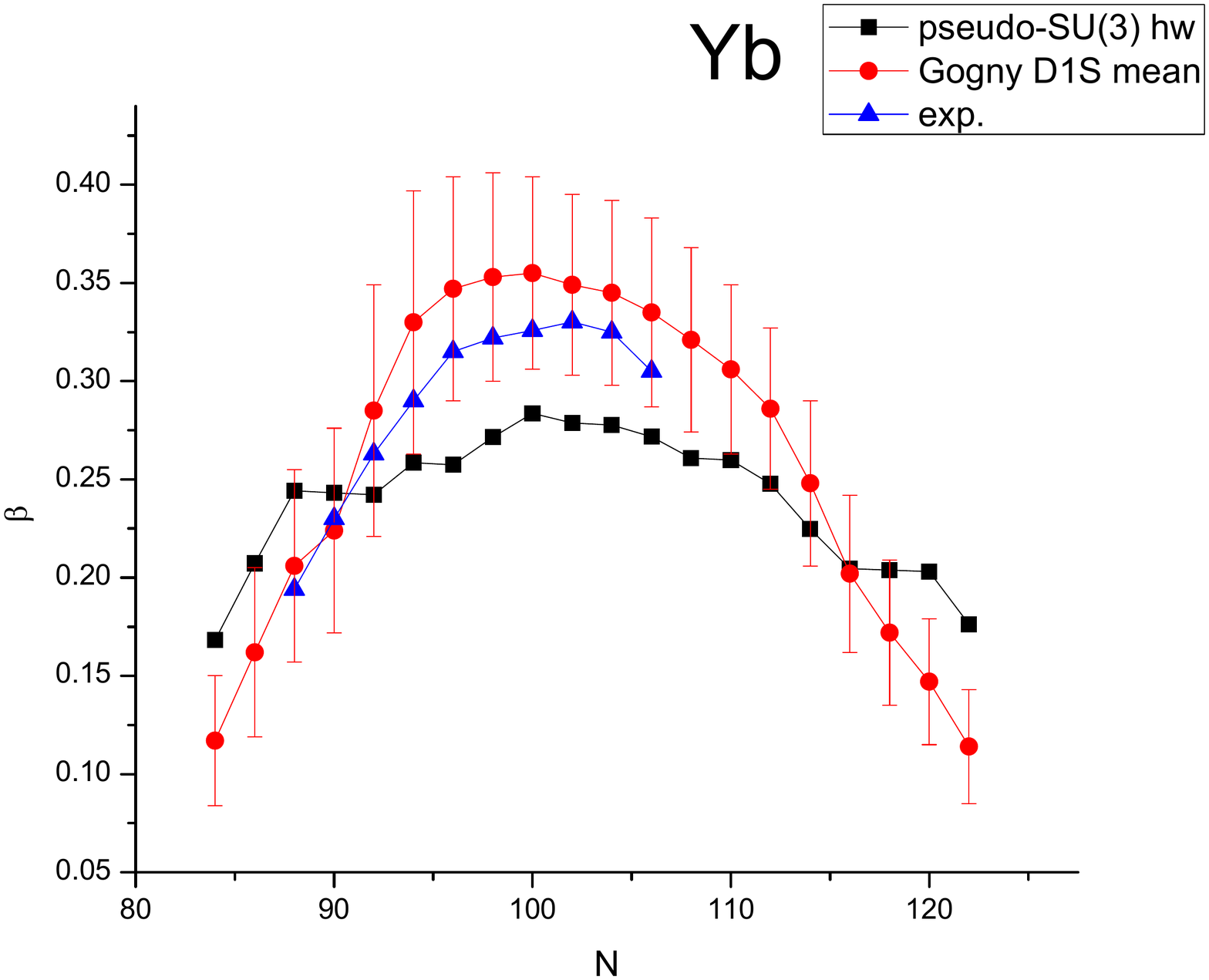}
\includegraphics[width=60mm]{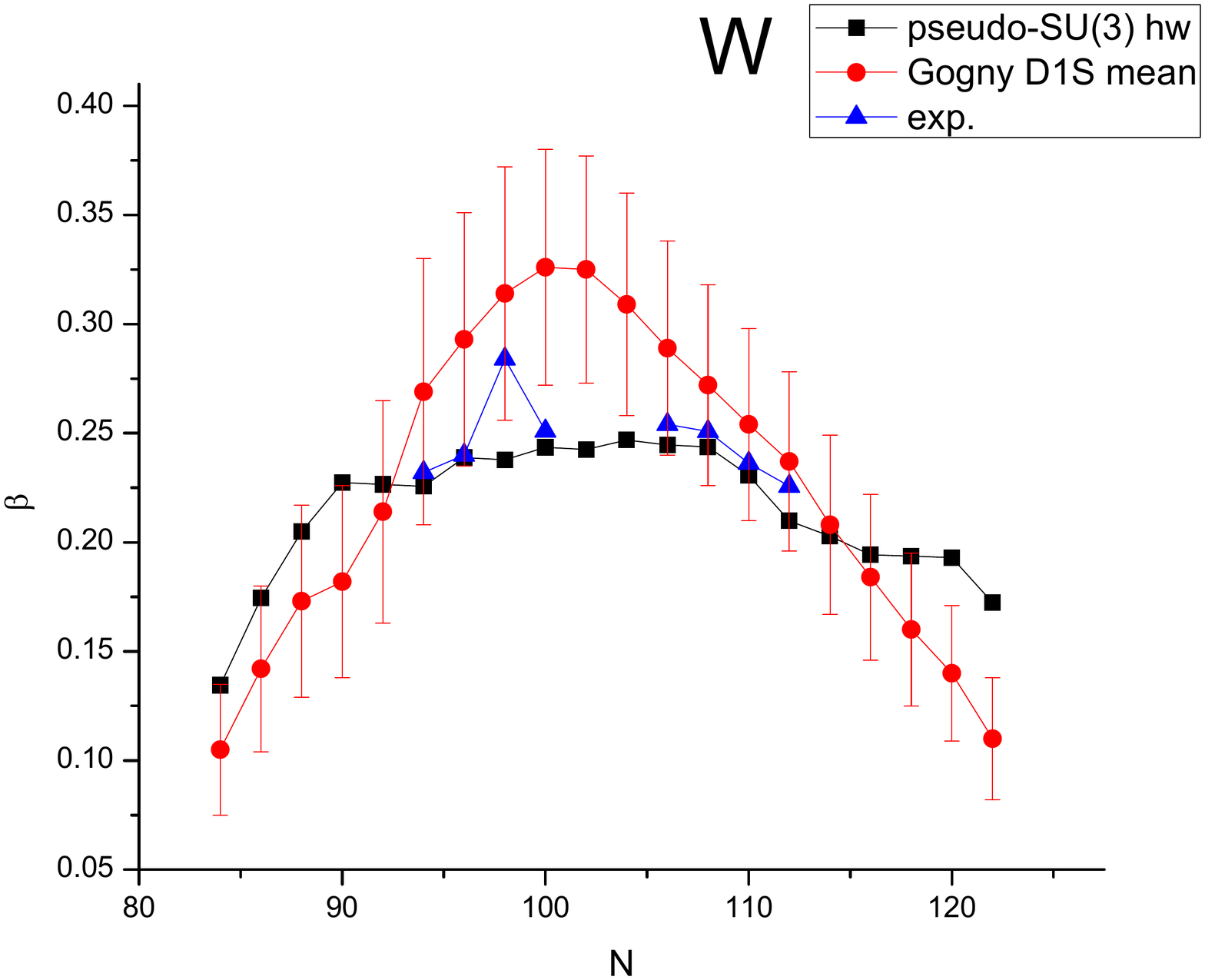}
\includegraphics[width=60mm]{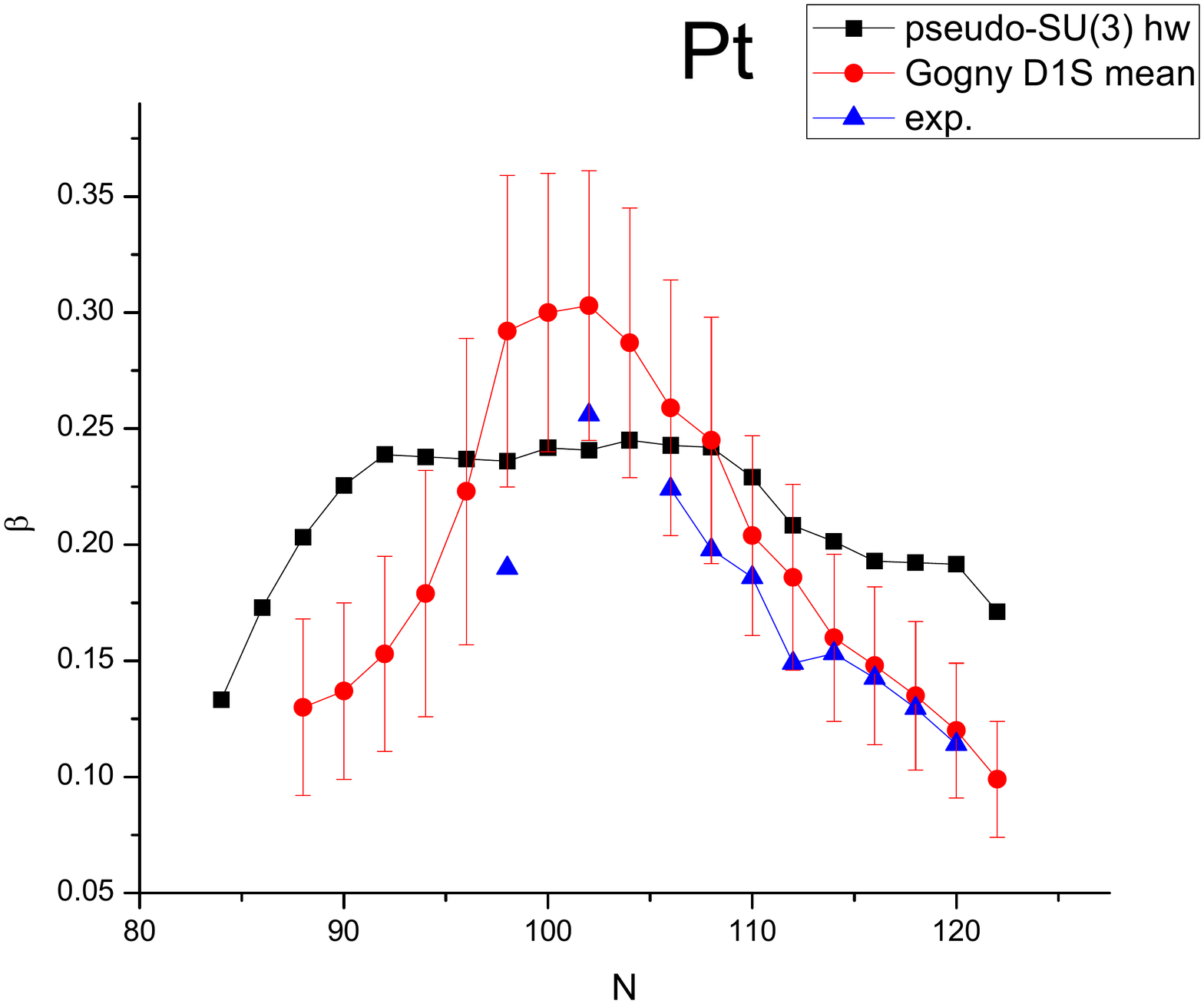}}

\caption{The pseudo-SU(3) hw predictions shown in Fig. \ref{2B} for the collective deformation variable $\beta$ for six series of isotopes in the rare earth region are compared to results by the D1S-Gogny interaction (Gogny D1S mean) \cite{Gogny}, as well as with empirical values (exp.) \cite{Pritychenko}. 
 See Section \ref{num} for further discussion.} 
\label{3B}
\end{figure*}

\begin{figure*}[htb]

{\includegraphics[width=60mm]{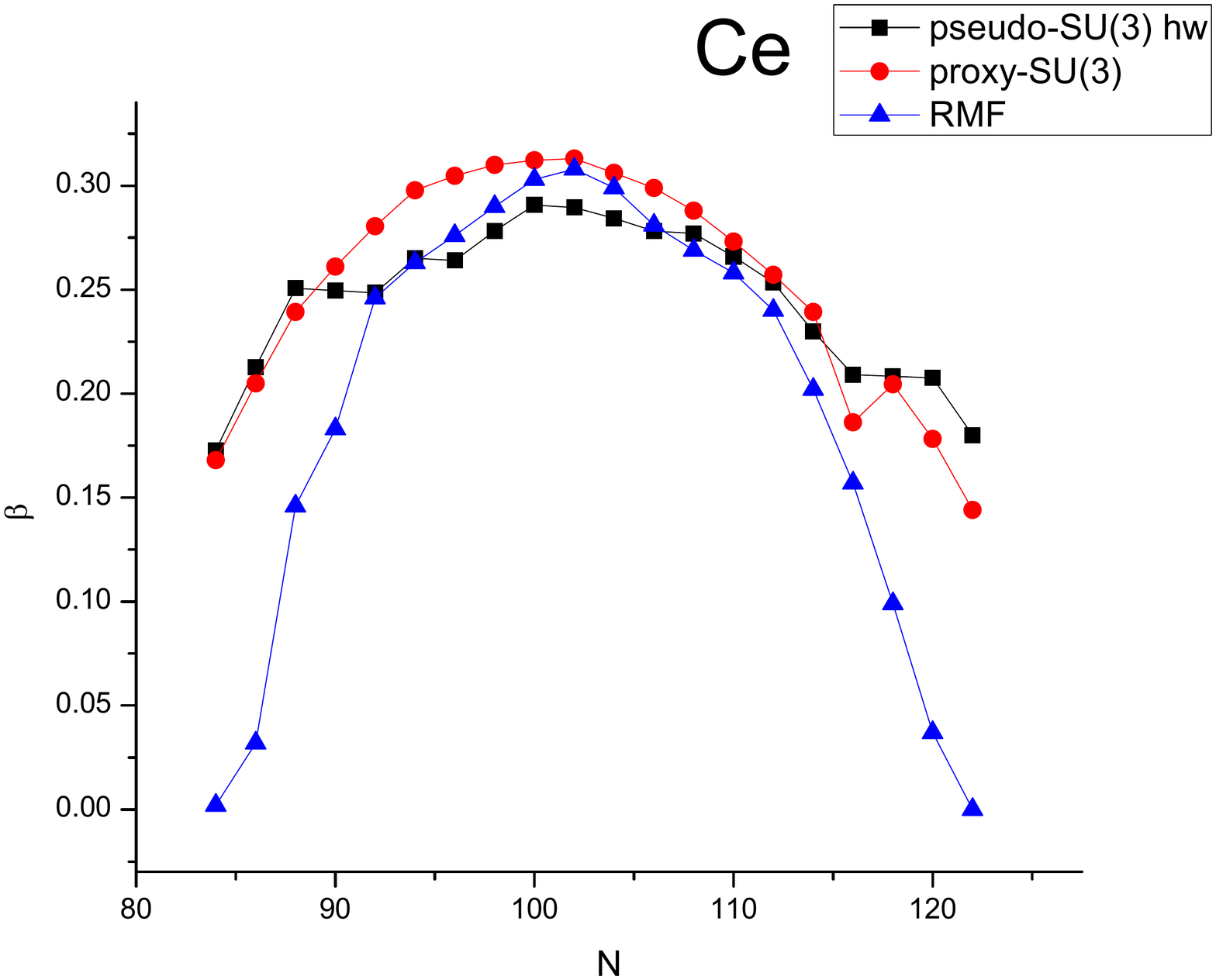}
\includegraphics[width=60mm]{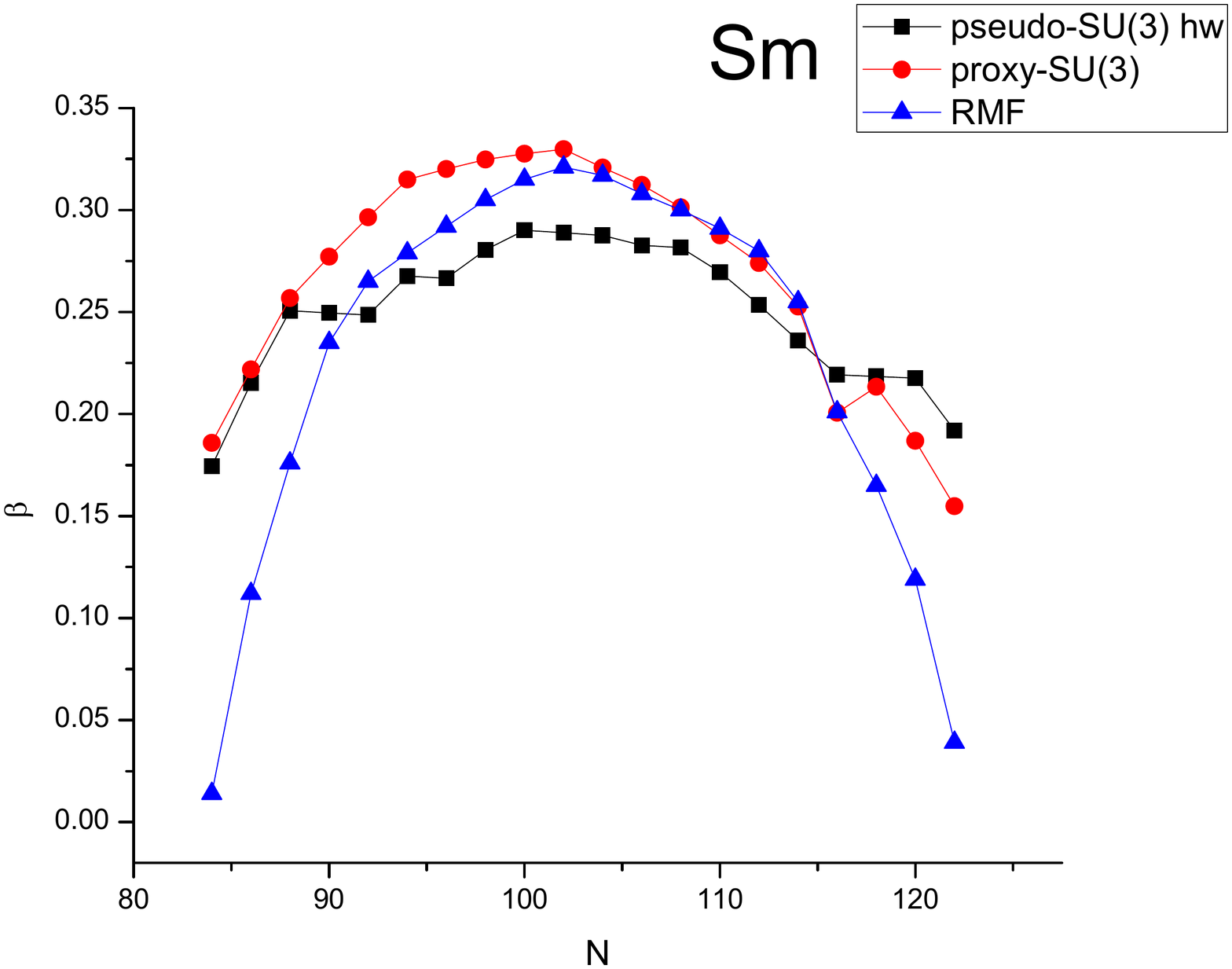}
\includegraphics[width=60mm]{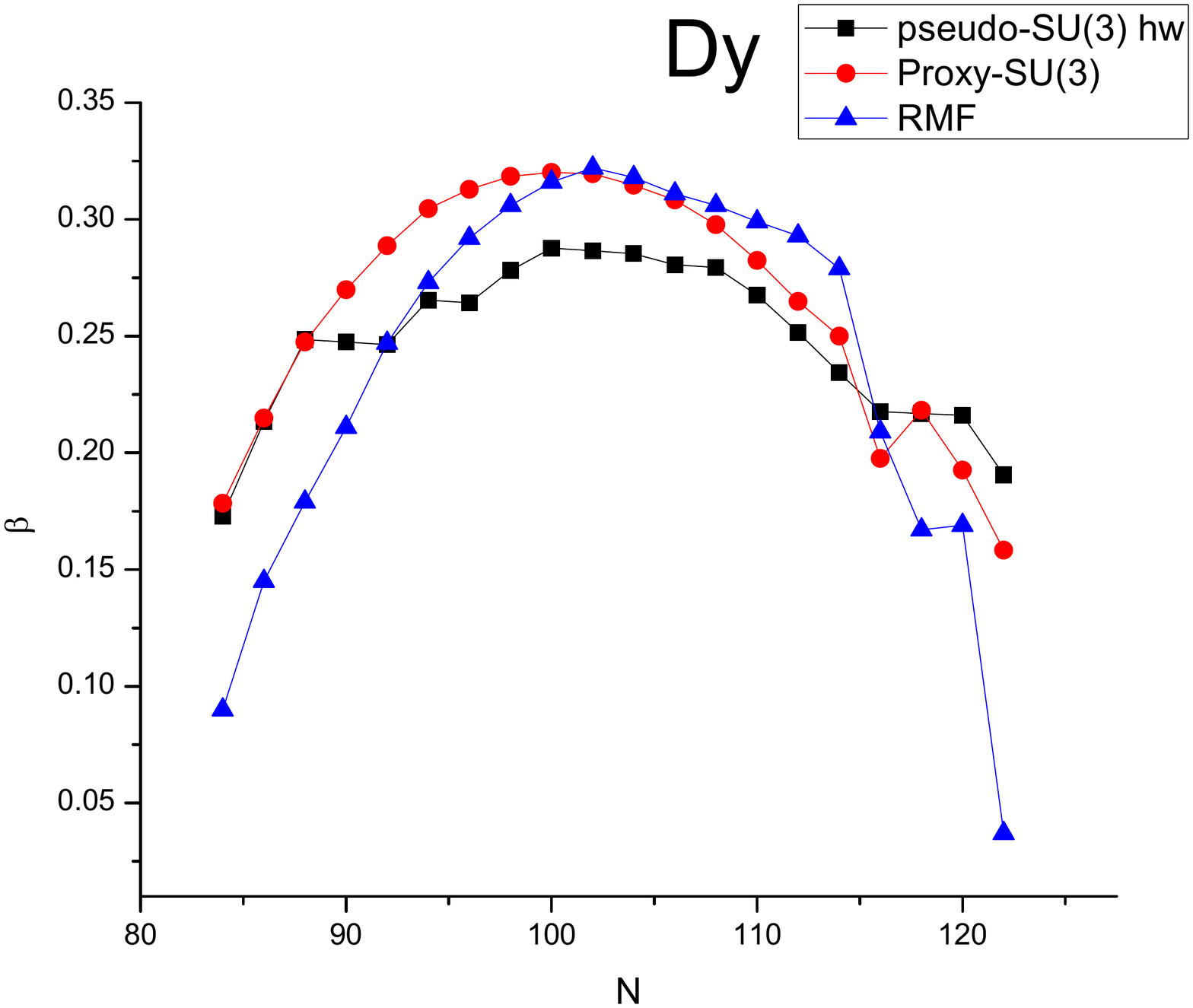}}
{\includegraphics[width=60mm]{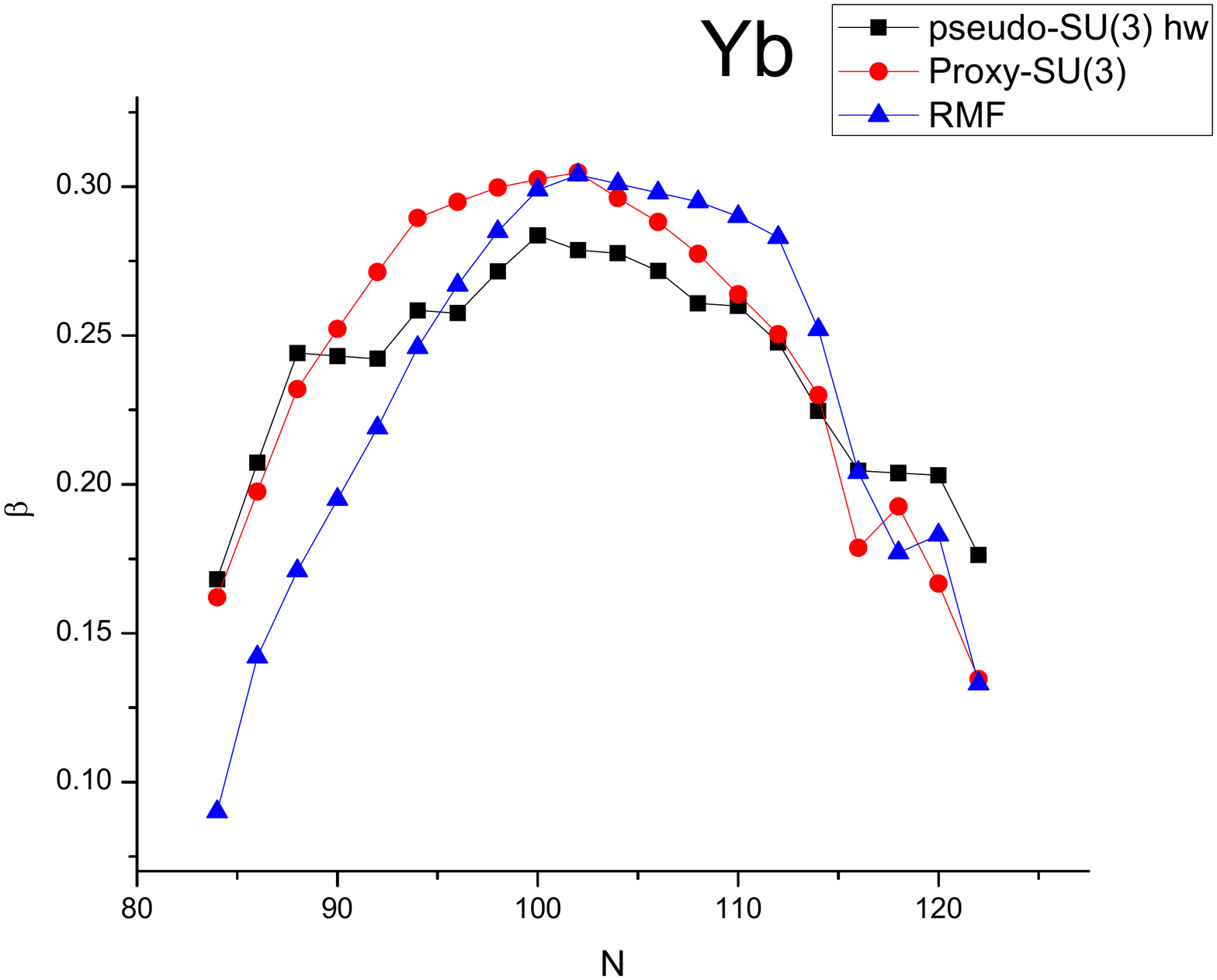}
\includegraphics[width=60mm]{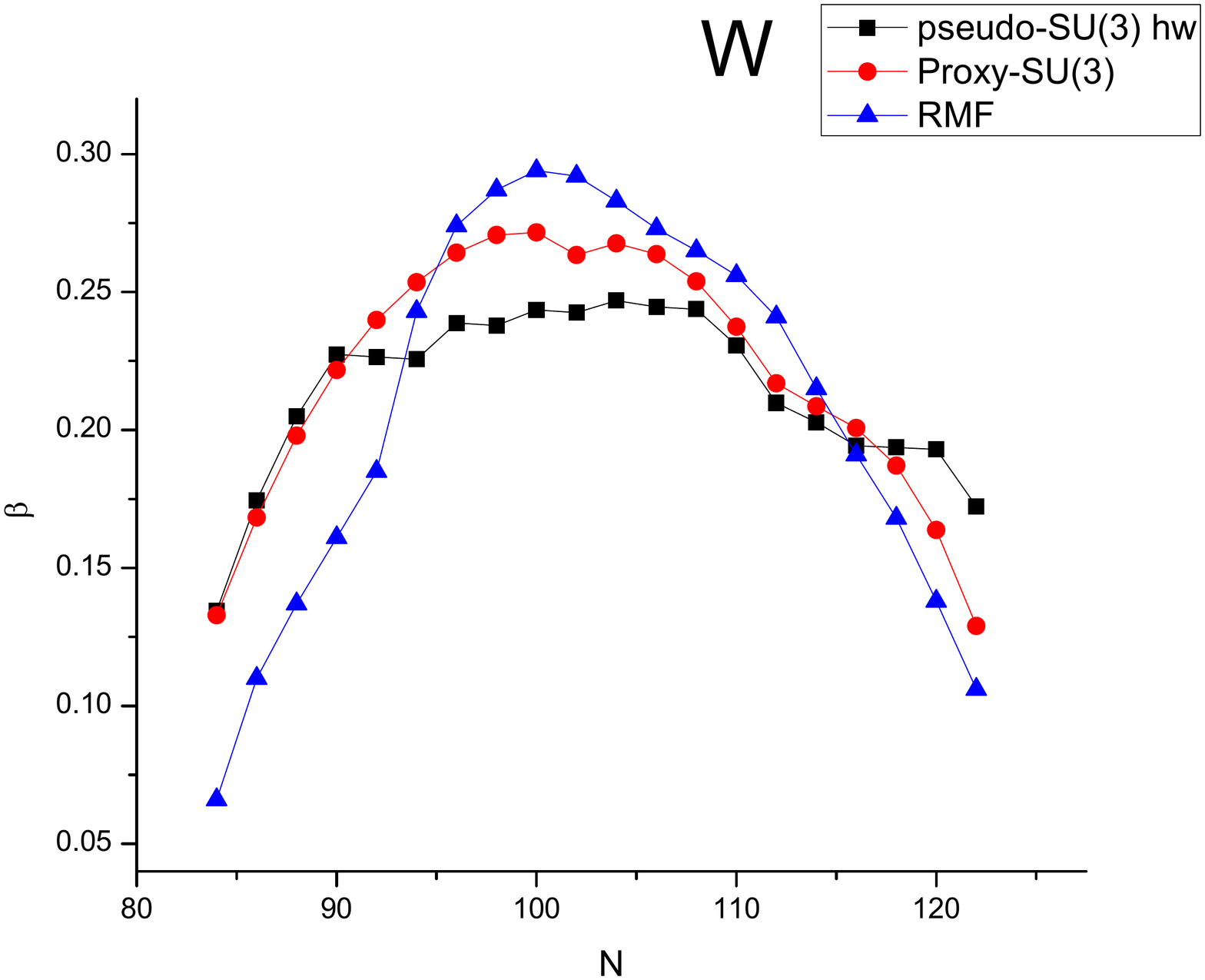}
\includegraphics[width=60mm]{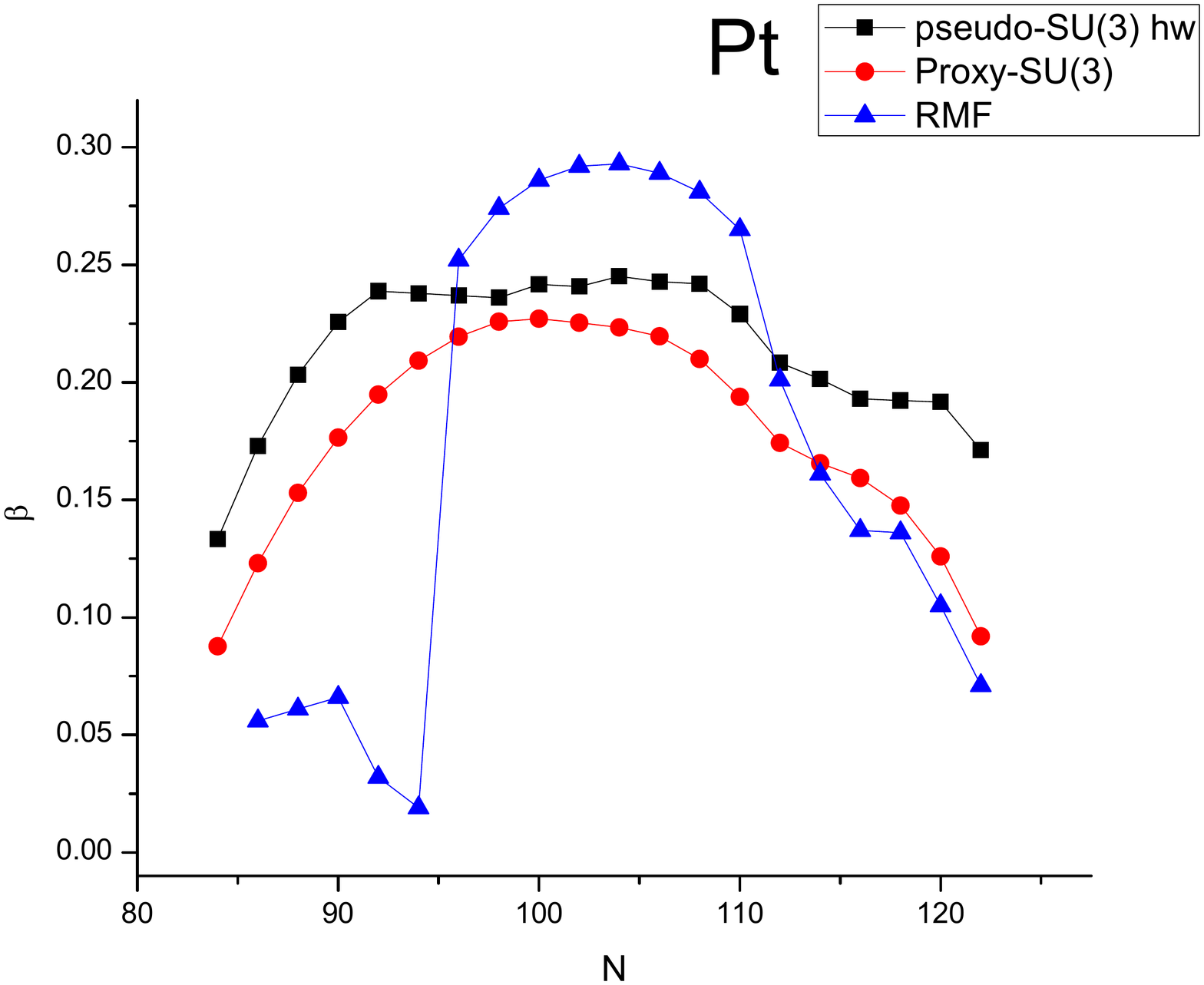}}

\caption{The pseudo-SU(3) hw predictions shown in Fig. \ref{2B} for the collective deformation variable $\beta$ for six series of isotopes in the rare earth region are compared to proxy-SU(3) results obtained as described in Ref. \cite{proxy2}, as well as to relativistic mean field theory (RMF) predictions\cite{Lalazissis}.
 See Section \ref{num} for further discussion.} 
\label{4B}
\end{figure*}

\begin{figure*}[htb]

{\includegraphics[width=60mm]{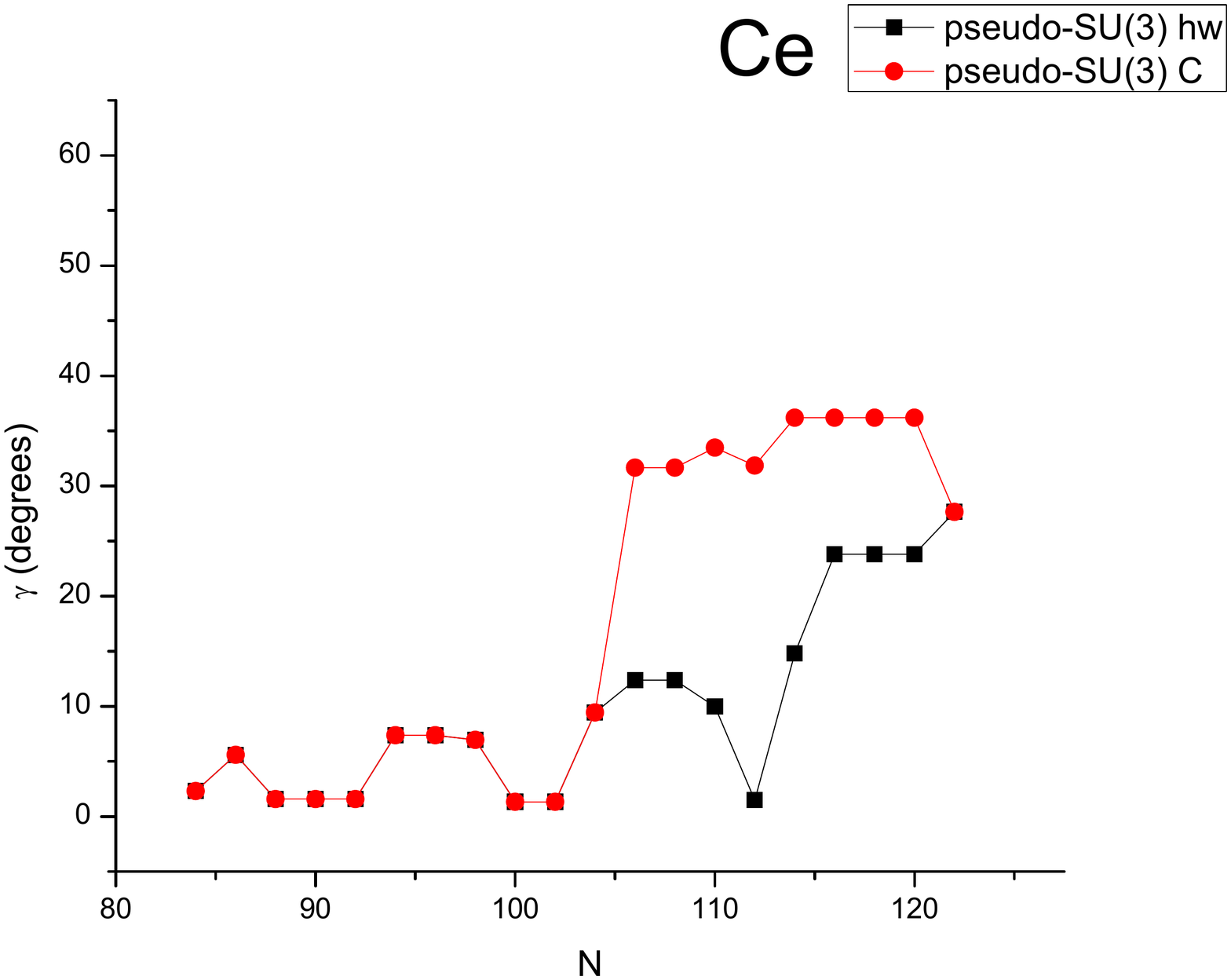}
\includegraphics[width=60mm]{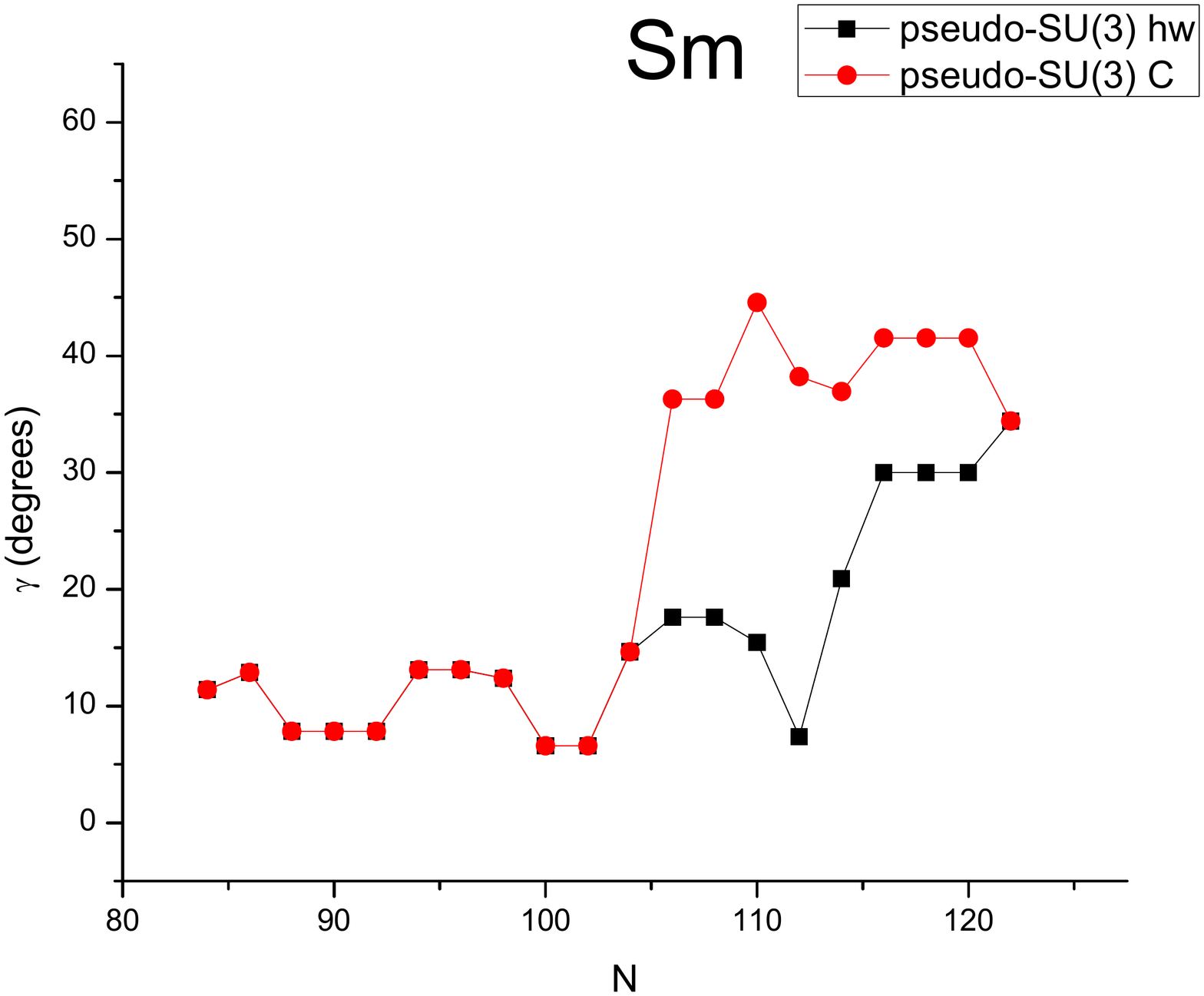}}
{\includegraphics[width=60mm]{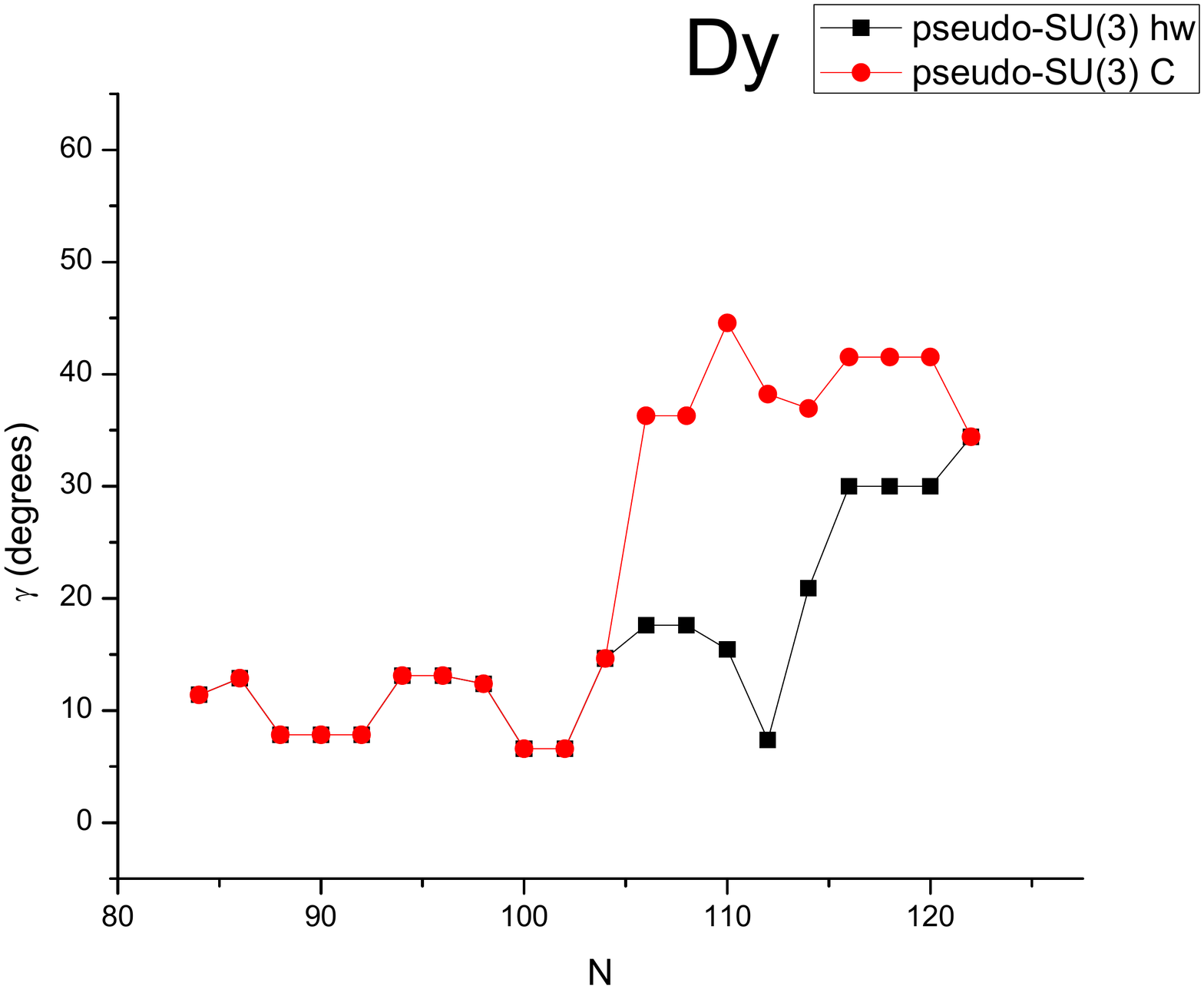}
\includegraphics[width=60mm]{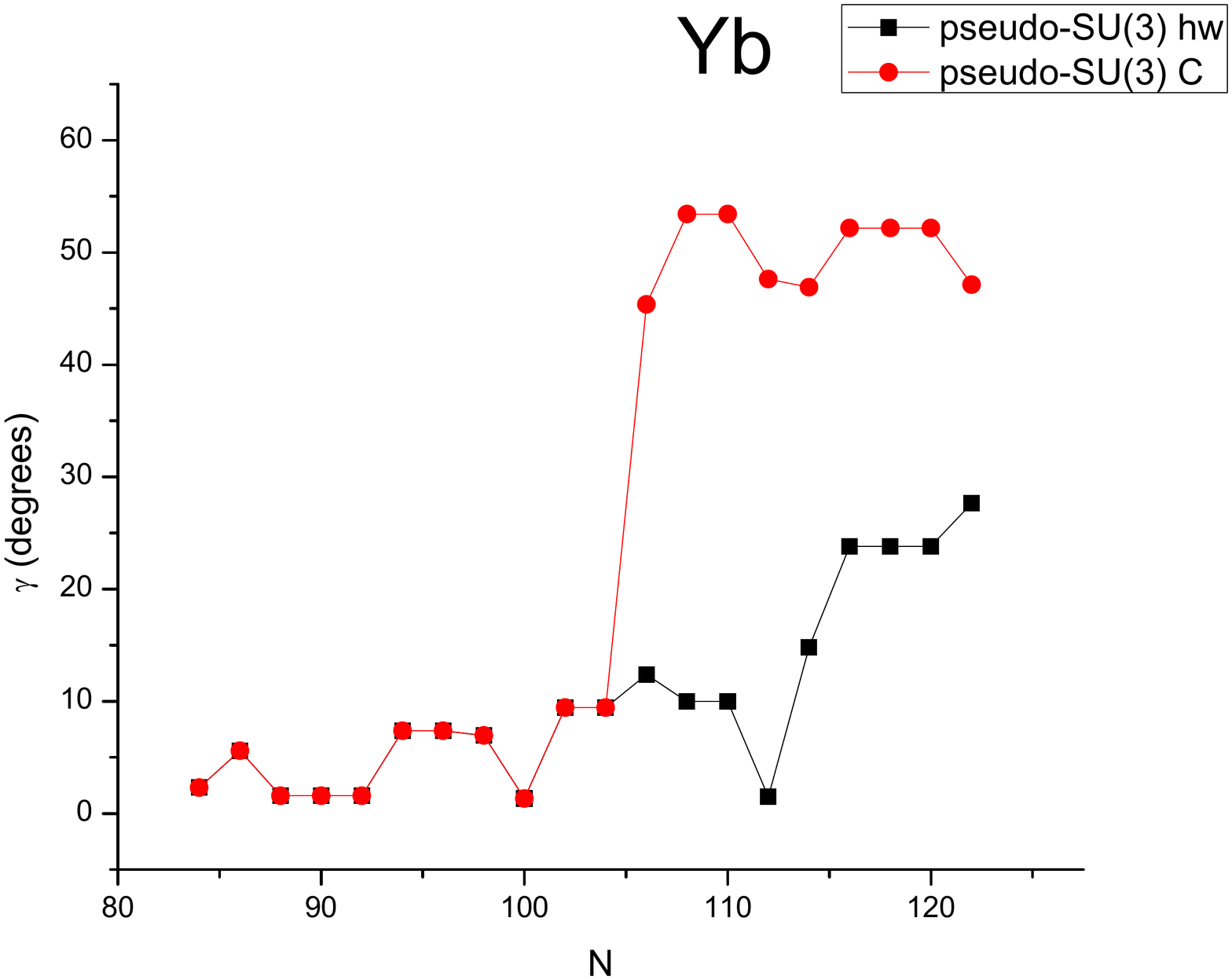}}
{\includegraphics[width=60mm]{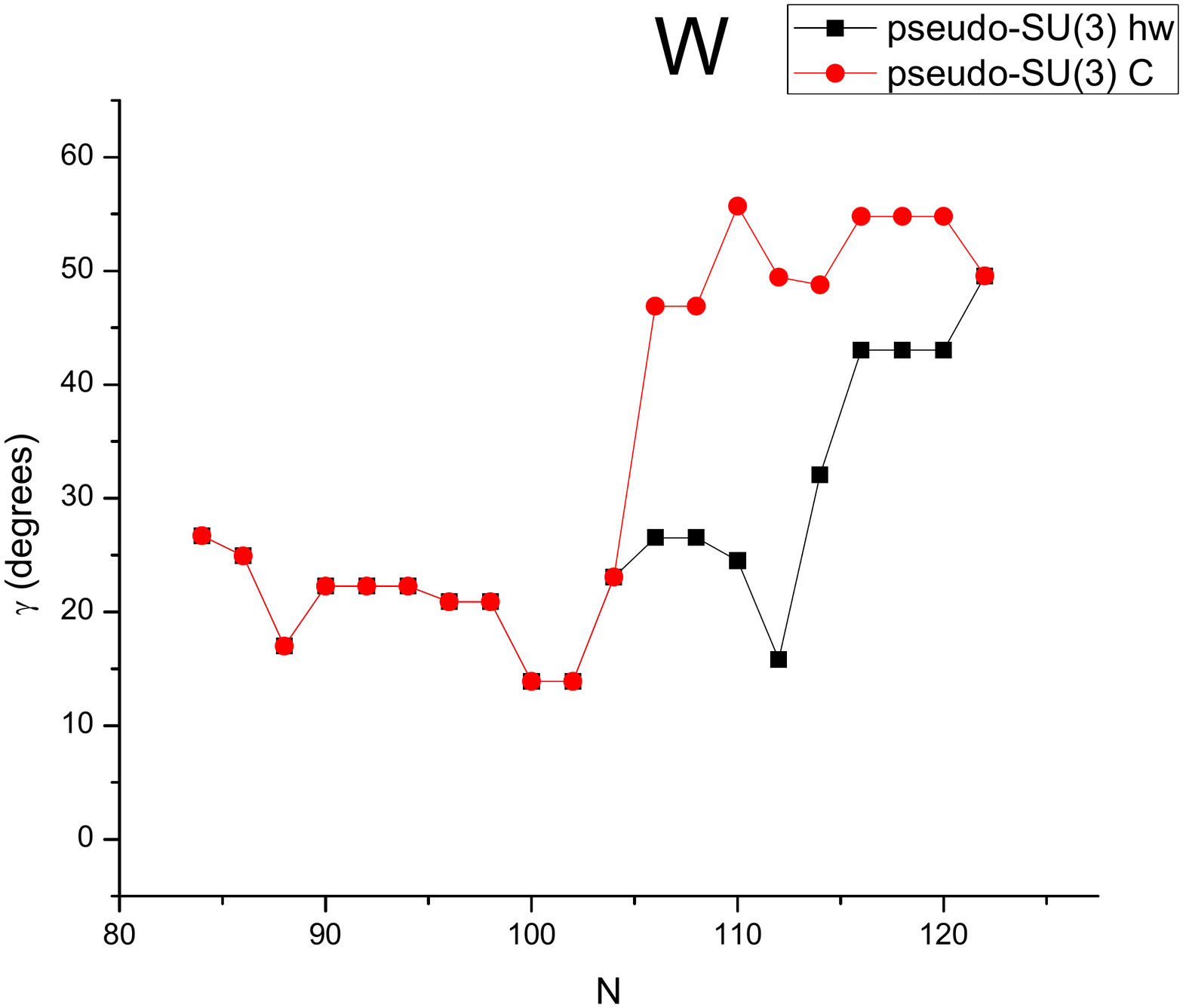}
\includegraphics[width=60mm]{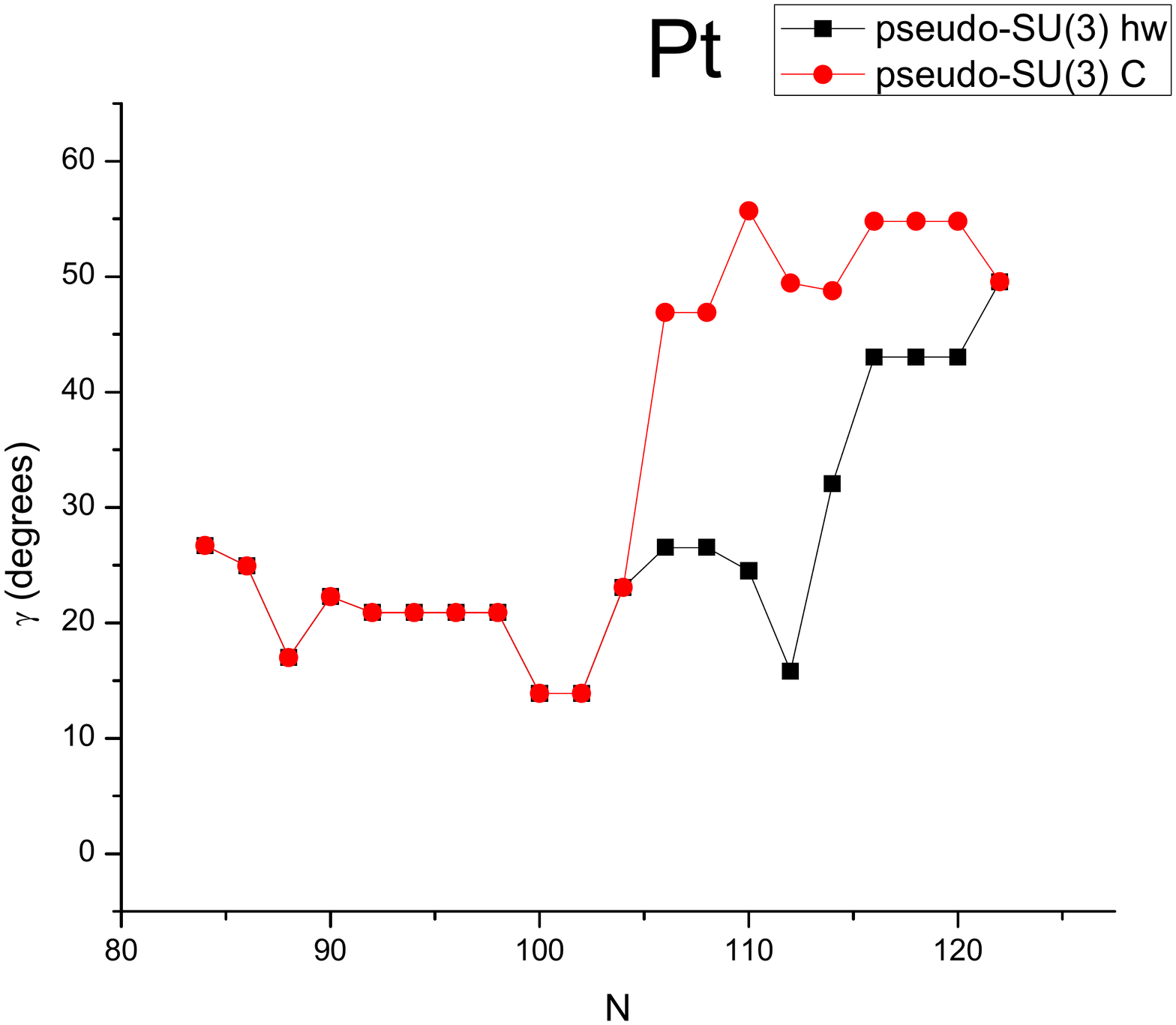}}

\caption{Pseudo-SU(3) predictions for the collective deformation variable $\gamma$ for six series of isotopes in the rare earth region. The predictions labeled by hw have been obtained using the highest weight irreps of SU(3), 
while those labeled by C have been obtained using the hC irreps of SU(3) having the highest eigenvalue 
of the second order Casimir operator of SU(3), $C_2^{SU(3)}$. 
 See Section \ref{num} for further discussion.}

\label{2G}
\end{figure*}

\begin{figure*}[htb]

\includegraphics[width=120mm]{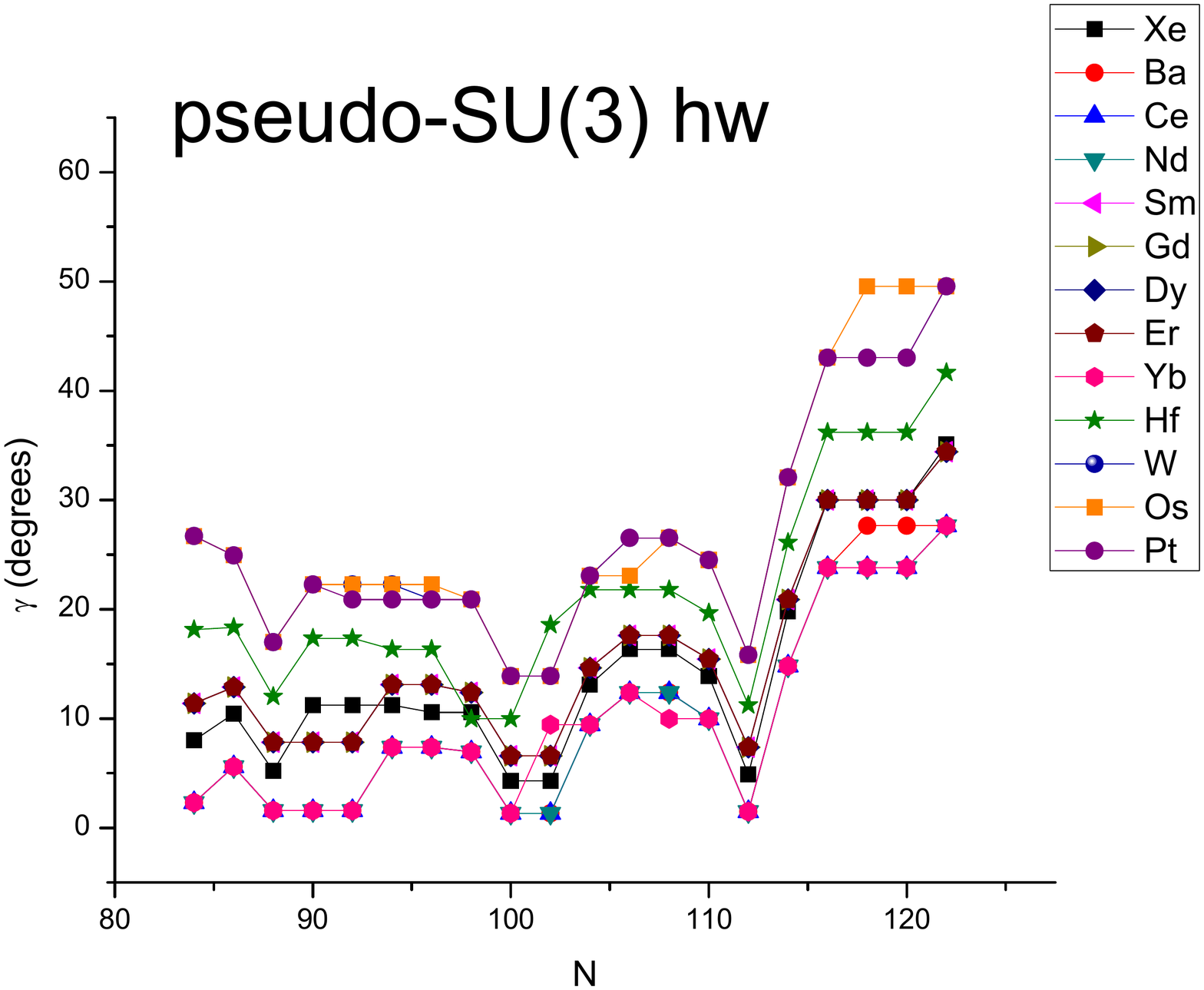} 
\includegraphics[width=120mm]{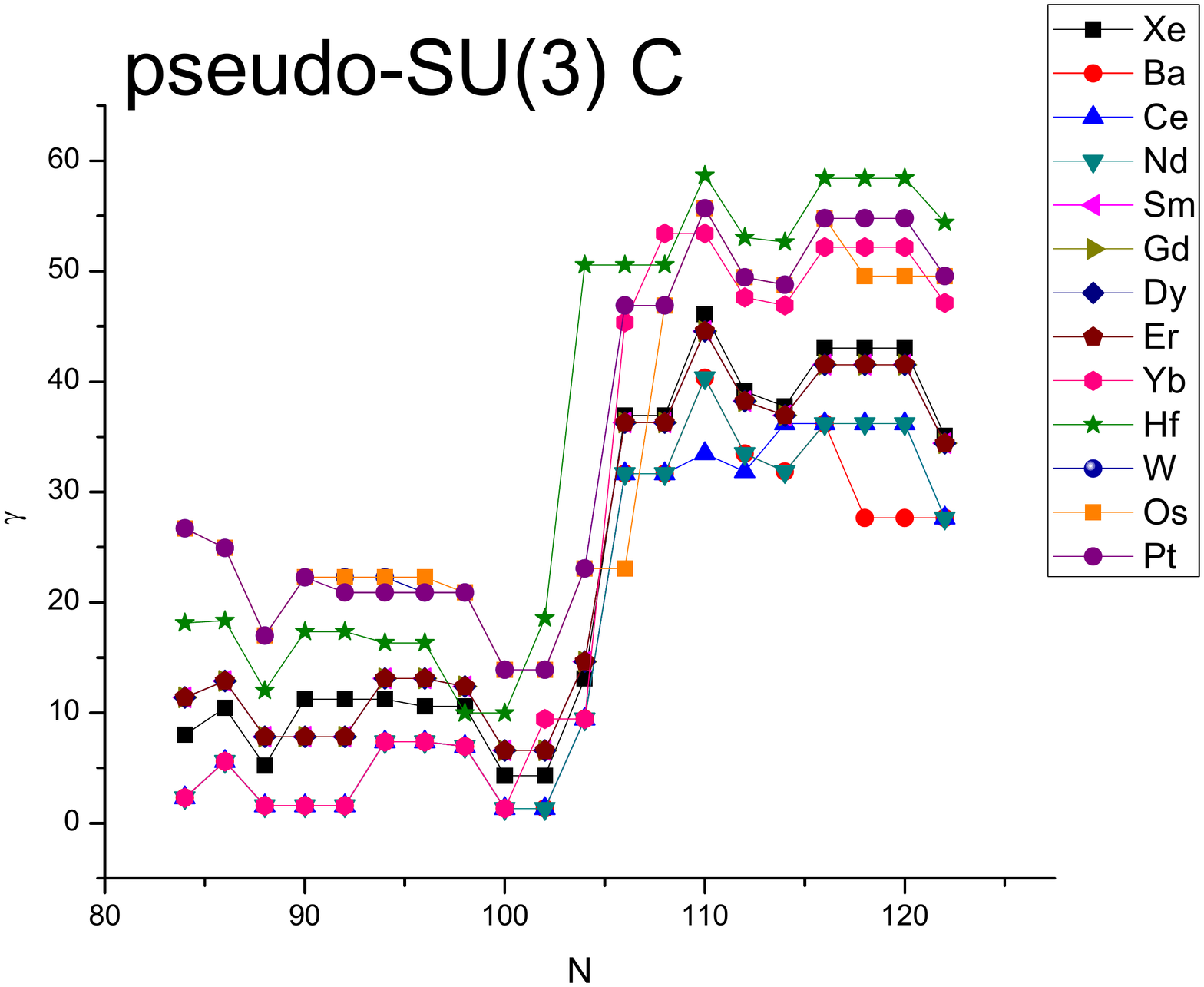}

\caption{Pseudo-SU(3) predictions for the collective deformation variable $\gamma$ for the Xe-Pt series of isotopes in the rare earth region. The predictions labeled by hw (top panel) have been obtained using the highest weight irreps of SU(3), 
while those labeled by C (bottom panel) have been obtained using the hC irreps of SU(3) having the highest eigenvalue 
of the second order Casimir operator of SU(3), $C_2^{SU(3)}$. Adopted from Ref. \cite{Rila19}.
 See Section \ref{num} for further discussion.} 
 
\label{G}
\end{figure*}

\begin{figure*}[htb]

{\includegraphics[width=60mm]{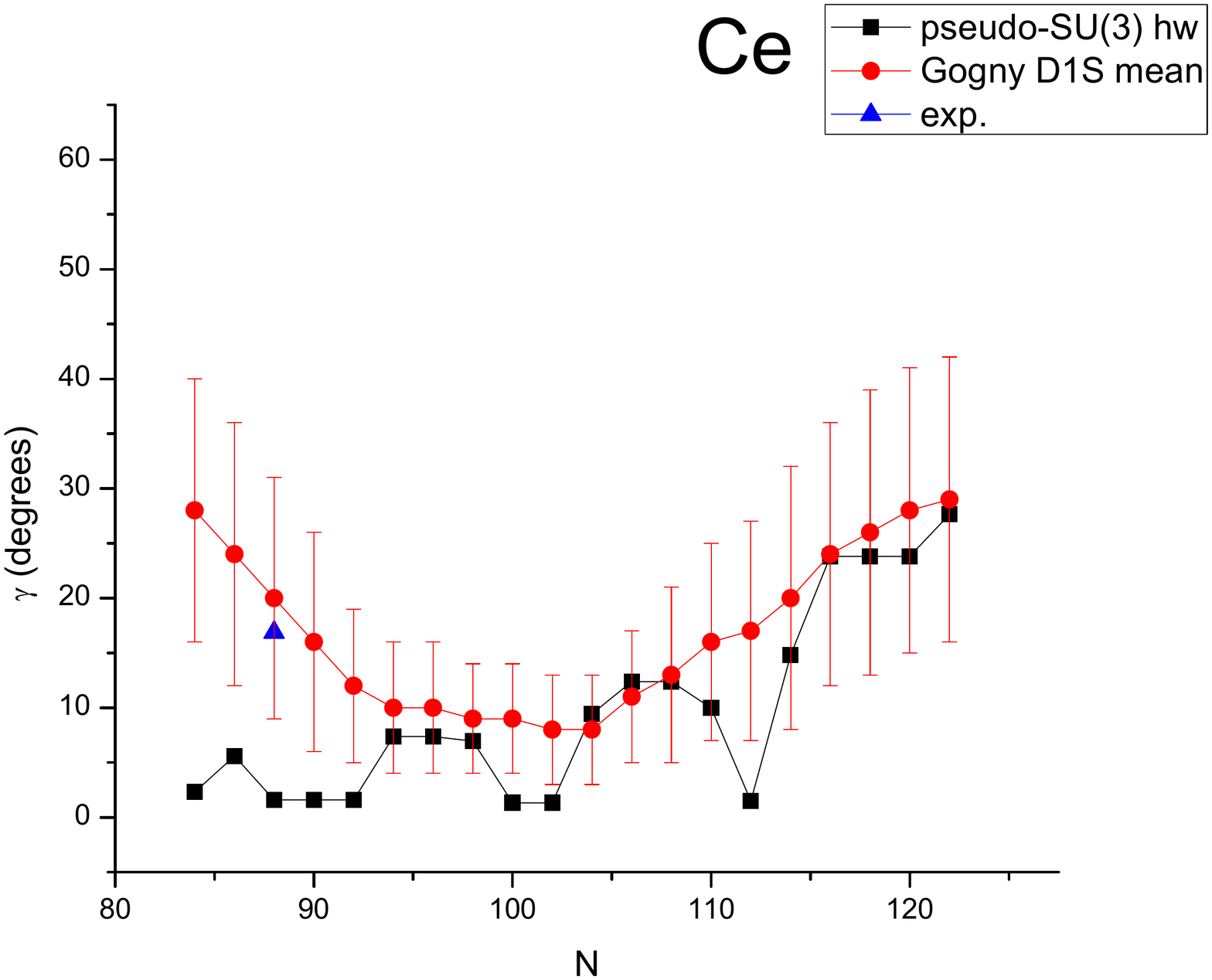}
\includegraphics[width=60mm]{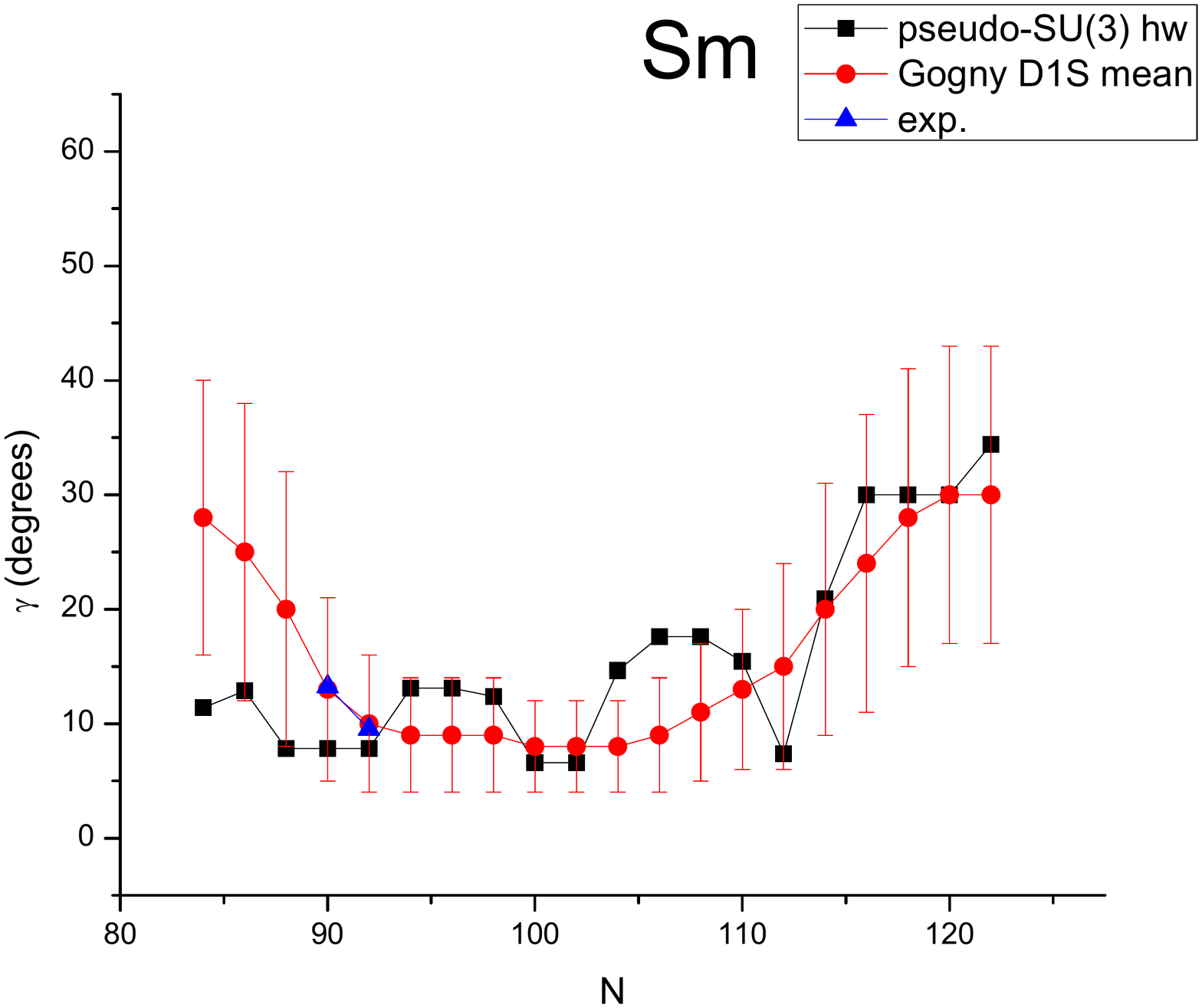}}
{\includegraphics[width=60mm]{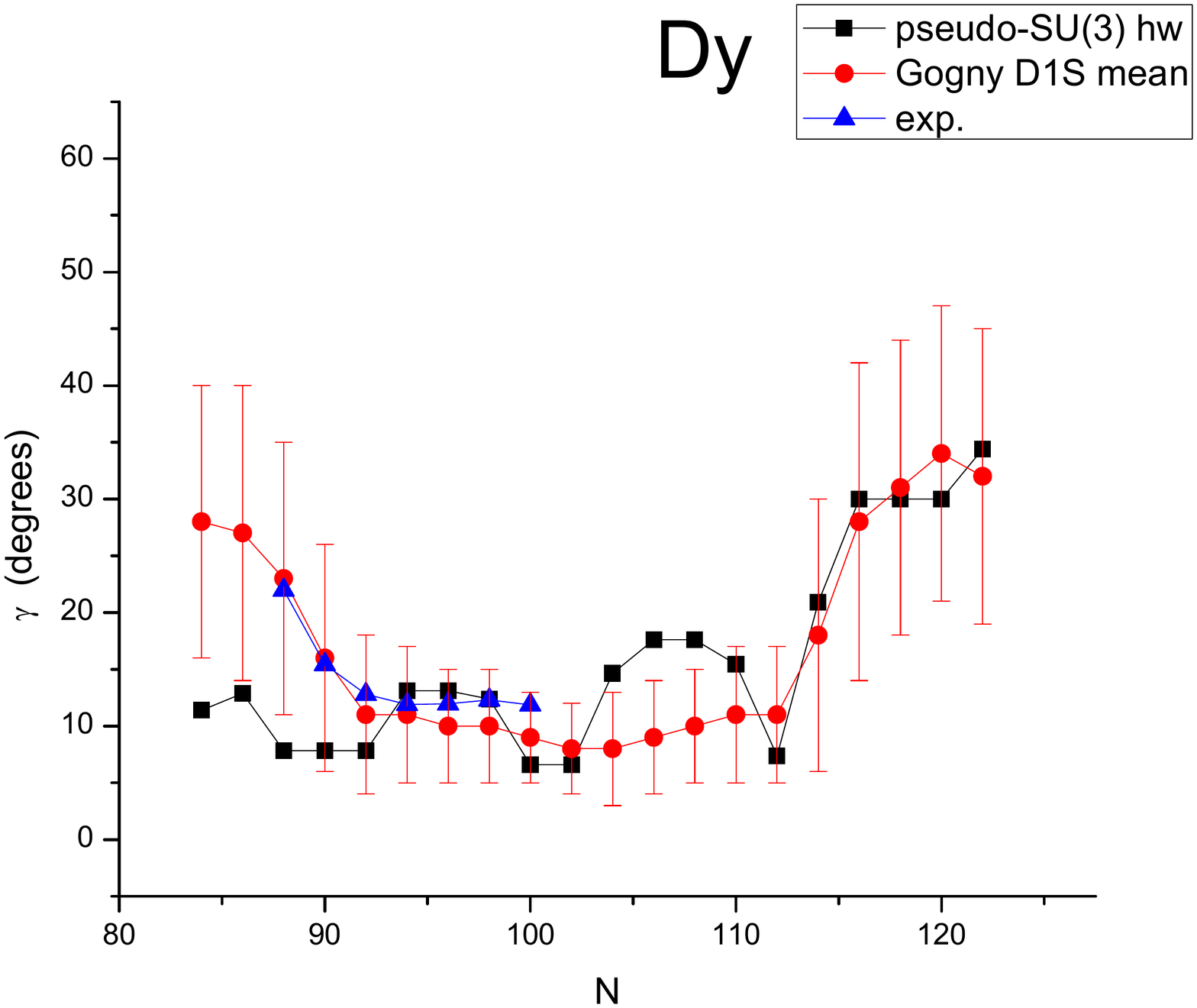}
\includegraphics[width=60mm]{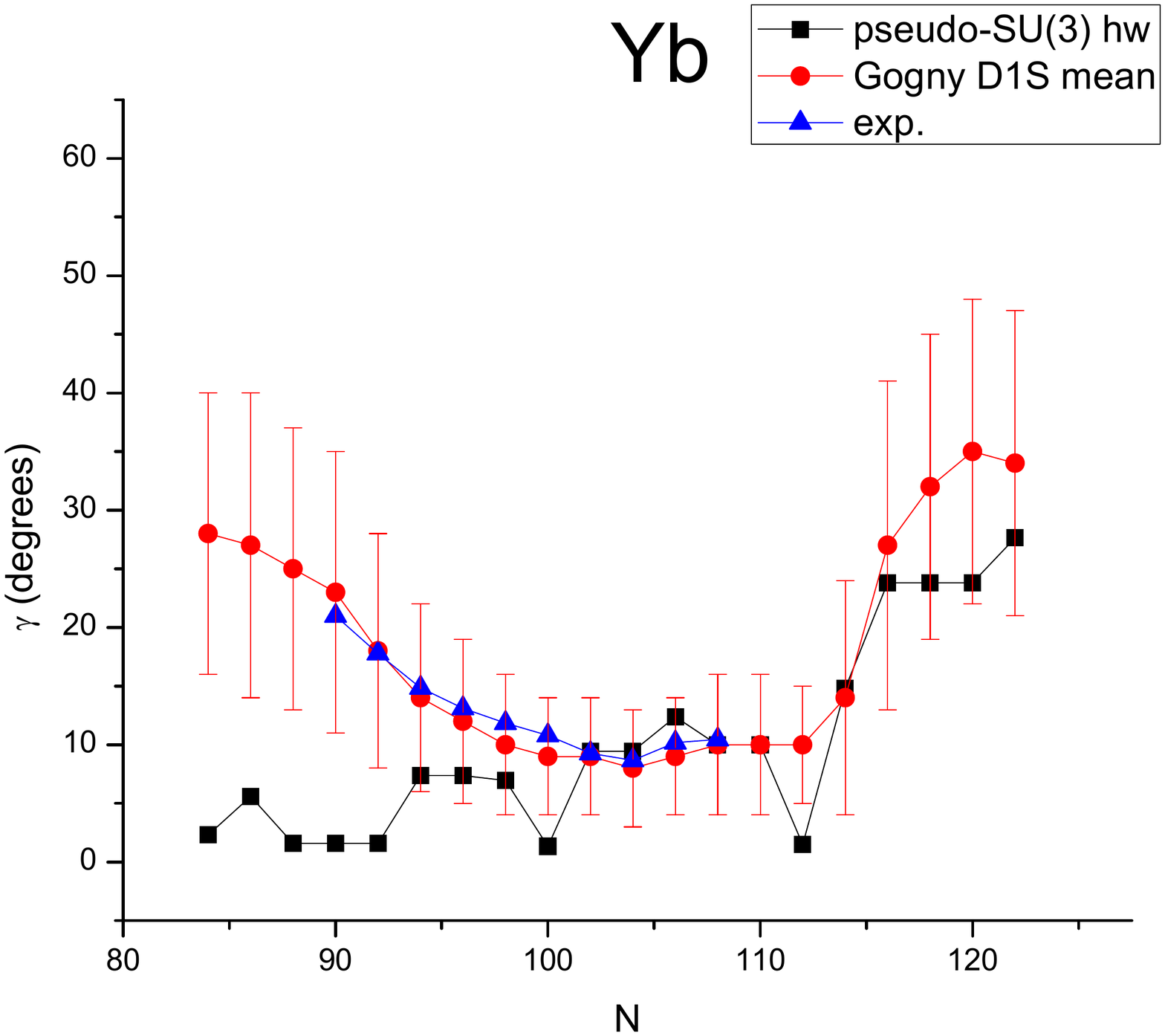}}
{\includegraphics[width=60mm]{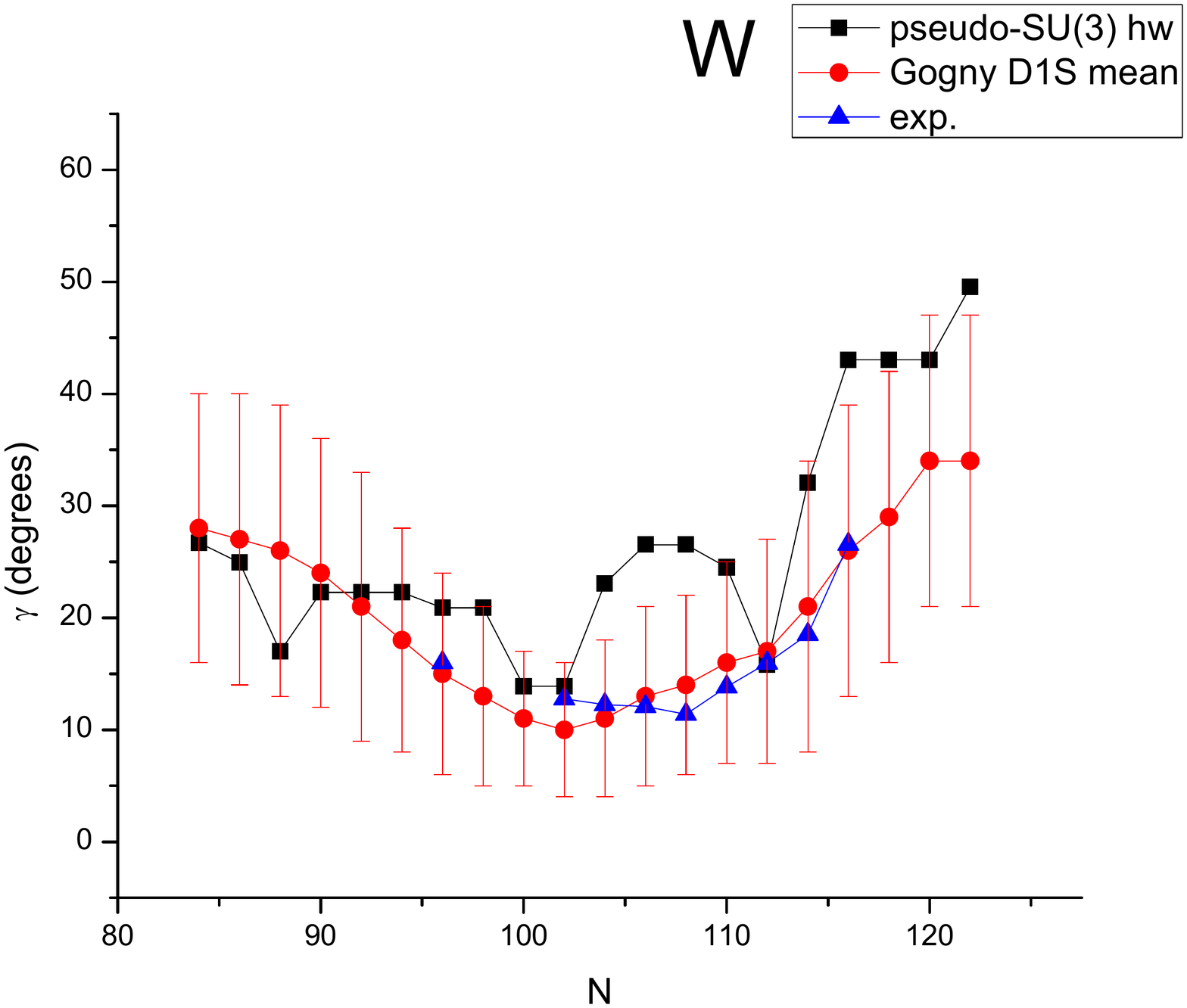}
\includegraphics[width=60mm]{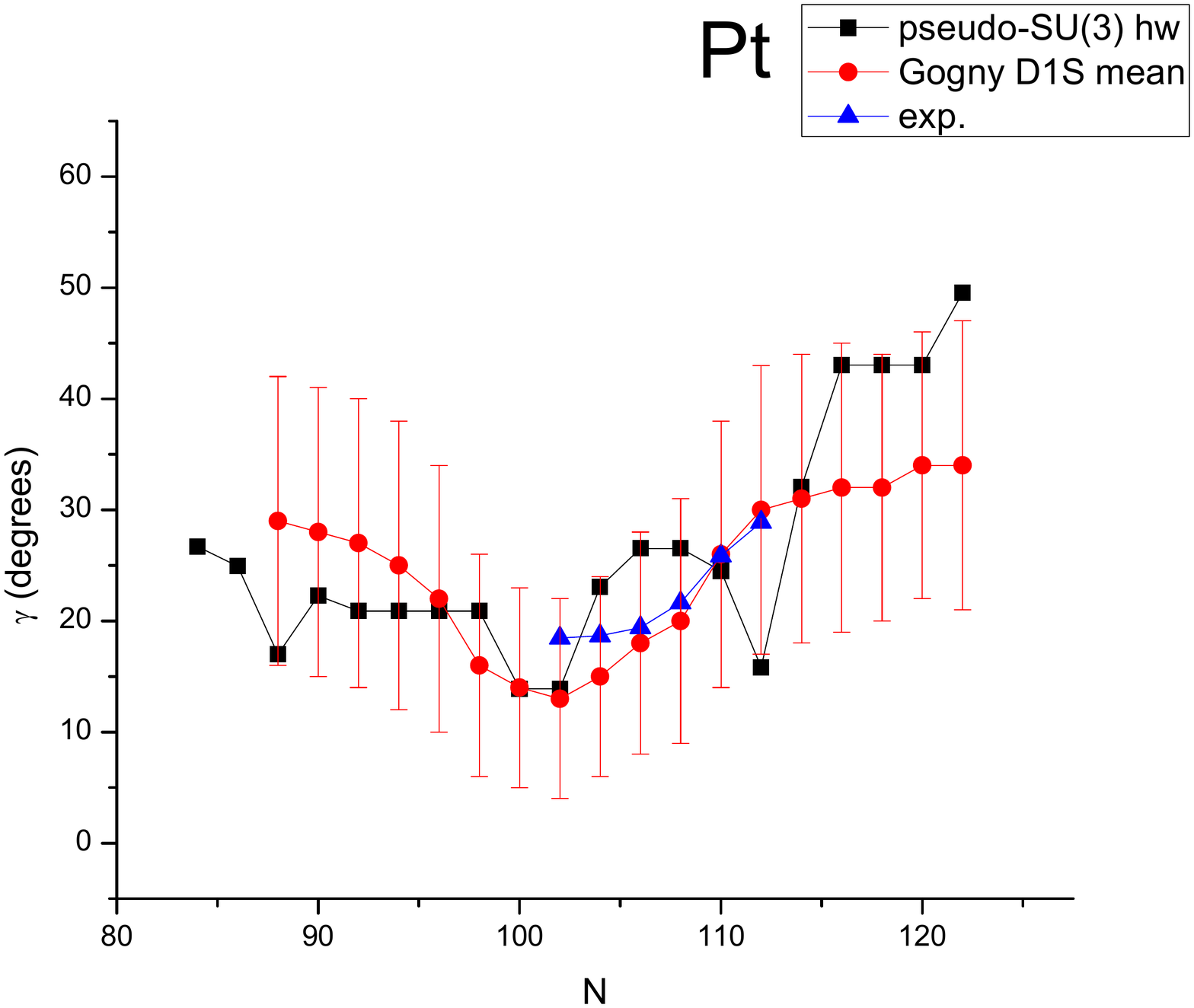}}

\caption{The pseudo-SU(3) hw predictions shown in Fig. \ref{2G} for the collective deformation variable $\gamma$ for six series of isotopes in the rare earth region are compared to results by the D1S-Gogny interaction (Gogny D1S mean) \cite{Gogny}, as well as with empirical values (exp.), calculated through Eq.
(\ref{gm}). 
 See Section \ref{num} for further discussion.} 

\label{3G}
\end{figure*}

\begin{figure*}[htb]

{\includegraphics[width=60mm]{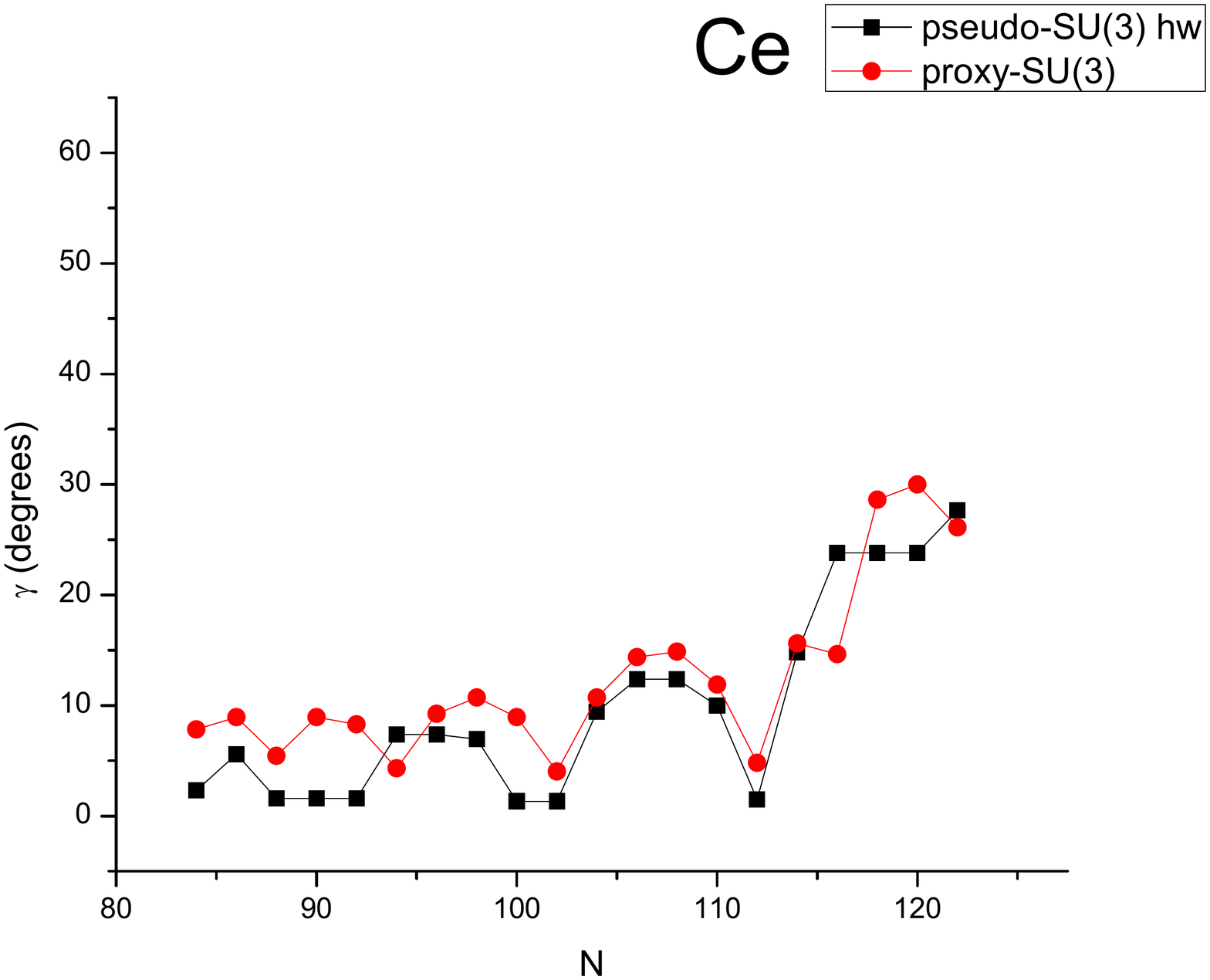}
\includegraphics[width=60mm]{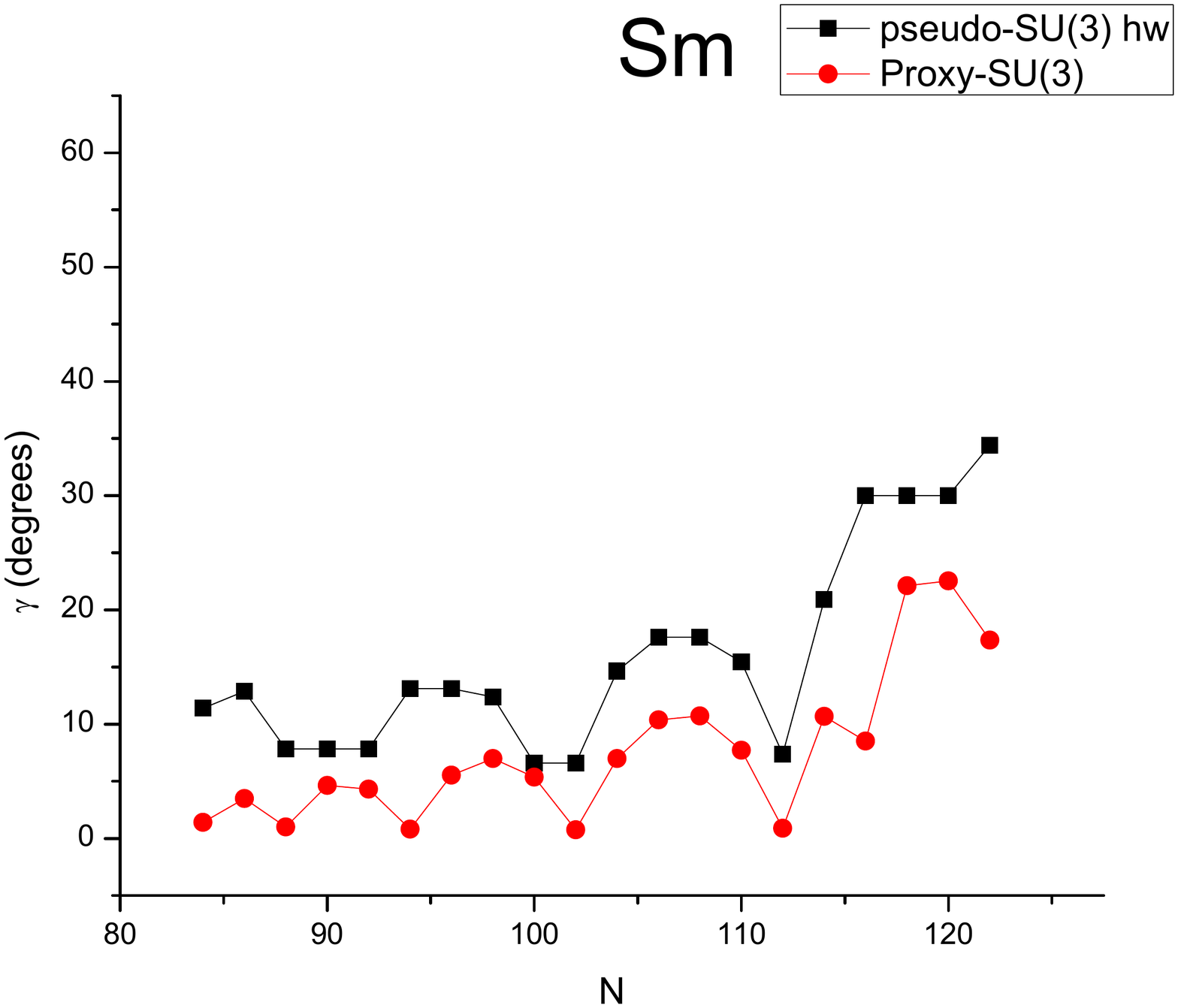}}
{\includegraphics[width=60mm]{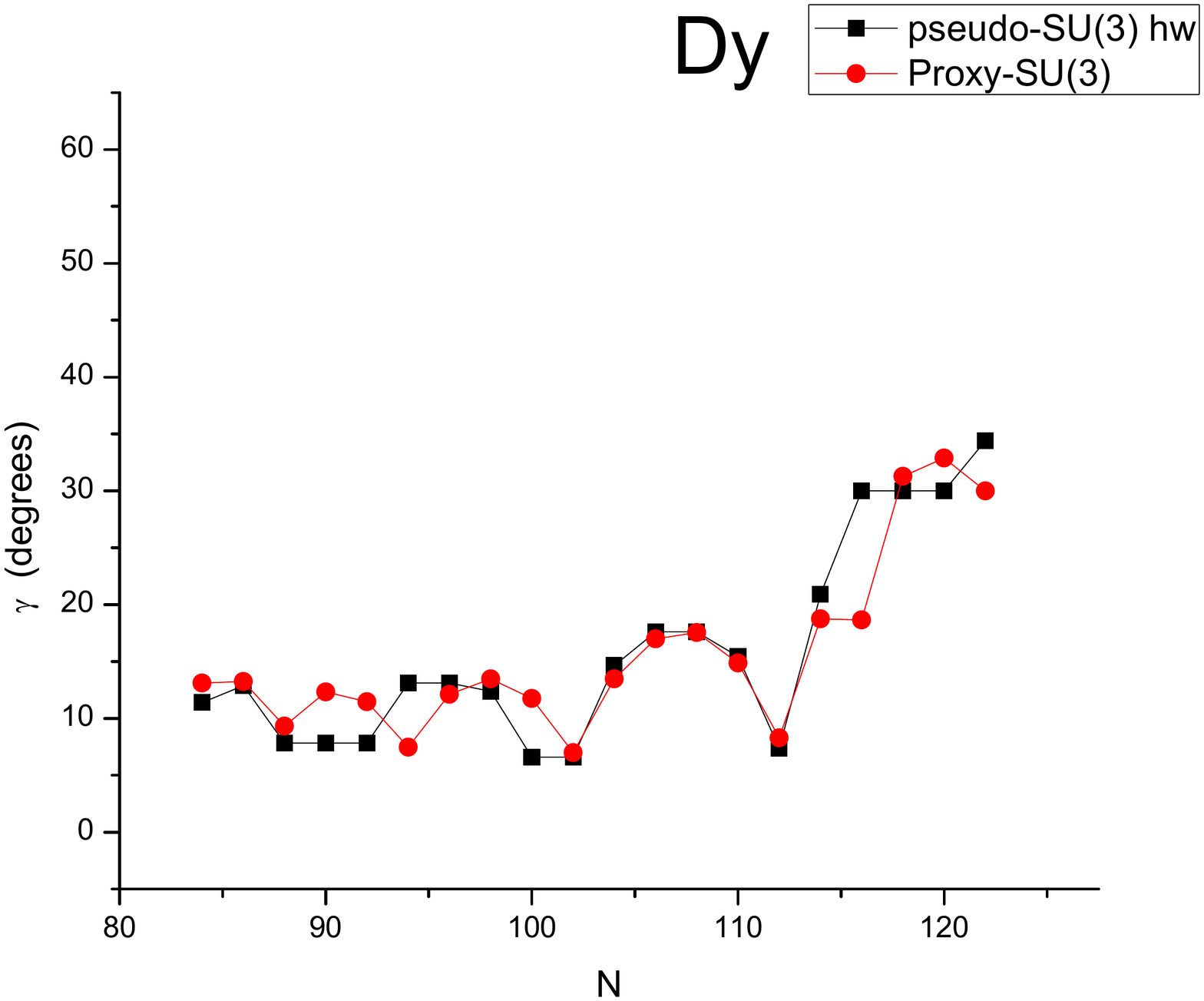}
\includegraphics[width=60mm]{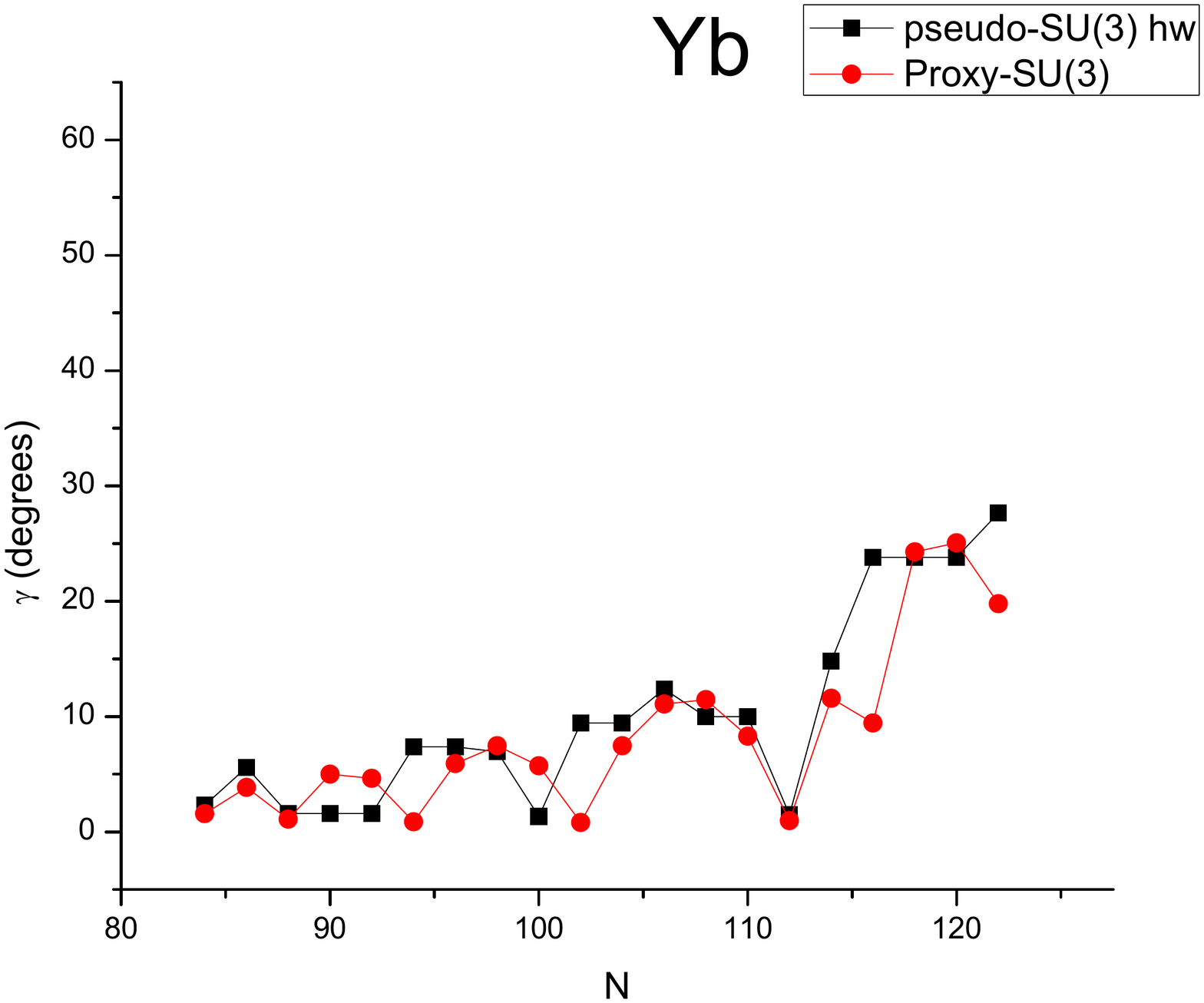}}
{\includegraphics[width=60mm]{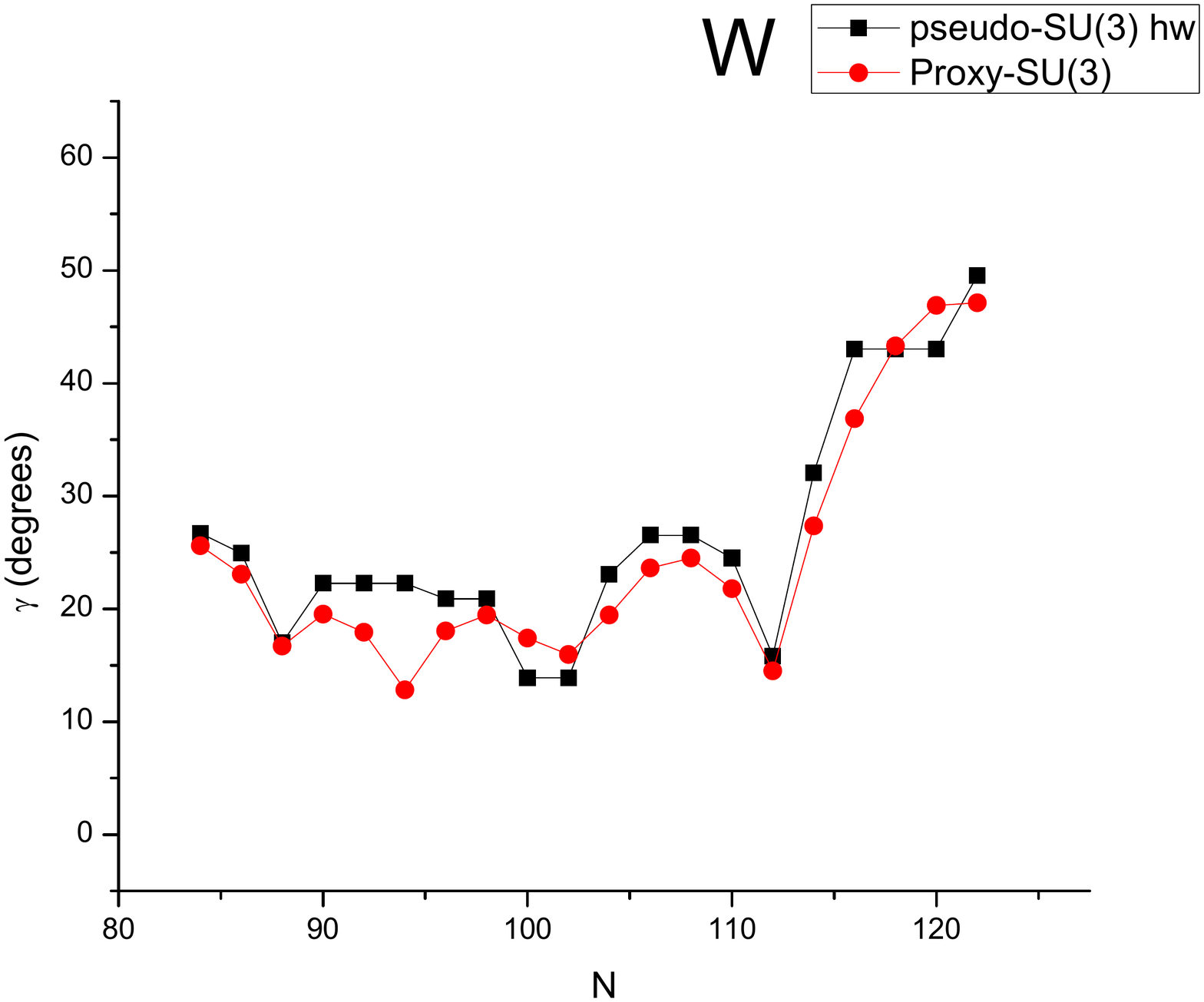}
\includegraphics[width=60mm]{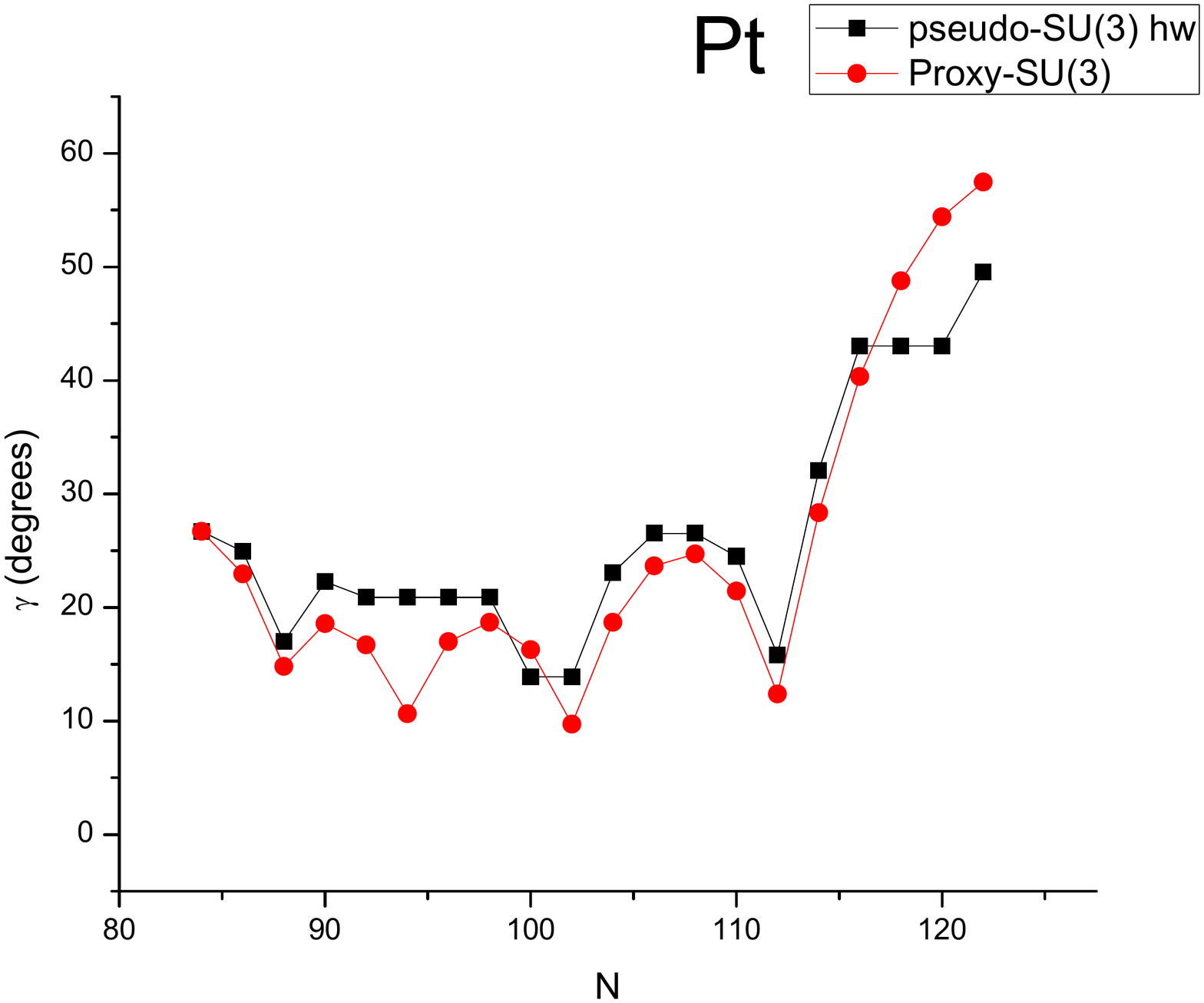}}

\caption{The pseudo-SU(3) hw predictions shown in Fig. \ref{2G} for the collective deformation variable $\gamma$ for six series of isotopes in the rare earth region are compared to proxy-SU(3) results obtained as described in Ref. \cite{proxy2}.
 See Section \ref{num} for further discussion.} 

\label{4G}
\end{figure*}

Numerical results for the collective deformation variables $\beta$ and $\gamma$ for several isotopic chains of rare earths are shown in Figs. (\ref{2B})-(\ref{4B}) and  Figs. (\ref{2G})-(\ref{4G}) respectively.

In Fig. (\ref{2B}) the influence of the choice of the hw irreps vs. the hC irreps on $\beta$ is shown. While in all cases the predictions for $\beta$ are identical 
up to the middle of the neutron shell, the hw predictions lie systematically lower than the hC predictions in the upper half of the shell. 

The influence of the choice of the hw irreps vs. the hC irreps is more dramatic on the $\gamma$ values, shown in Fig.(\ref{2G}). 
Again the predictions are identical up to the midshell, whereas beyond it they diverge dramatically. The hC predictions immediately jump above the $\gamma=30^{\rm o}$ border between prolate and oblate shapes, thus giving a clear sign for a prolate to oblate shape transition in the middle of the shell, which is not seen experimentally, while the hw predictions cross the prolate to oblate border only in the case of the W and Pt isotopic chains. 

The nearly horizontal segments appearing in both figures correspond to the gradual filling of intruder 
 neutron orbitals, as seen in Table 1. When a series of intruder neutron orbitals gets gradually filled,
the number of the neutrons in the ``remaining'' orbitals remains constant and therefore the relevant SU(3) irrep remains unchanged.  

An interesting detail can be noticed in Fig. (\ref{2G}). In the Ce, Sm, Dy, W, and Pt isotopic chains the hw and C curves meet again at N=122, while in the Yb isotopic chain they do not. This is due to the fact that 
in the corresponding harmonic oscillator shells the hw and hC irreps, which are identical up to the midshell but follow different paths above it, do meet again at the end of the shell for the last 4 particles in each shell, as seen in Table 2. From Table 1 we see that for all isotopic chains considered, at $N=122$ one has  26 valence neutrons of normal parity, which means 4 particles below the end of the shell, where the hw and hC irreps become again identical. Therefore for $N=122$ the hw and hC neutron irreps will be identical in all chains of isotopes. However, one has also to take into account the proton irreps. From Table 1 we see that 
the proton irreps will be identical in the Xe to Er isotopic chains, which possess up to 10 valence protons 
of normal parity, i.e. lie in the first half of the U(10) shell. They will also be identical in the W, Os, and Pt isotopic chains, which lie within the last four particles below the full U(10) shell. 
Therefore the only isotopic chains in which the hw and hC irreps for protons differ are the Yb and Hf chains.
As a consequence, the $\gamma$ values in Fig. (\ref{2G}) do not converge at $N=122$ for the Yb isotopes, because the hw and hC proton irreps are different, although the hw and hC neutron irreps are identical. 

The effects mentioned above can also be seen in Figs. (\ref{B}) and (\ref{G}), in which the $\beta$ and $\gamma$ predictions for all isotopic chains from Xe to Pt are summarized. In Fig. (\ref{B}) it is clear that 
both the hw and C choices give identical predictions for $\beta$ up to midshell, while above midshell 
the hw predictions  for $\beta$ lie systematically lower than the C predictions for all isotopic chains shown. In Fig. (\ref{G}) it is clear that 
both the hw and C choices give identical predictions for $\gamma$ up to midshell. Above midshell
the C predictions jump up to oblate values immediately after midshell, while the hw predictions cross the prolate to oblate border of $\gamma=30^{\rm{o}}$ much later, at $N=114$-116, and this only happens for the Hf, W, Os, and Pt isotopic chains, in agreement to existing experimental evidence, which has been extensively reviewed in Ref. \cite{proxy2} and needs not to be repeated here. 

In what follows attention is focused on the hw predictions of pseudo-SU(3). 

In Fig. (\ref{3B}) the hw pseudo-SU(3) predictions for $\beta$ are compared to predictions by the D1S Gogny interaction \cite{Gogny} and to the available experimental values \cite{Pritychenko}. In all isotopic chains the hw pseudo-SU(3) predictions lie lower than the D1S Gogny values near midshell, while they are higher than the D1S Gogny predictions near the beginning and near the end of the neutron shell. The same comments 
can be made when comparing the hw pseudo-SU(3) predictions to the data in the Ce, Sm, Dy, Yb, and Pt isotopic chains, while in the W isotopic chain the agreement of the hw pseudo-SU(3) predictions to the data is impressive. 

In Fig. (\ref{3G}) the hw pseudo-SU(3) predictions for $\gamma$ are compared to predictions by the D1S Gogny interaction \cite{Gogny} and to the available empirical values obtained as described in Sec. 3. In several cases the hw pseudo-SU(3) predictions lie within the error bars of the D1S Gogny predictions, which are in very good agreement with the empirical values. 

 In Fig. (\ref{4B}) the predictions for $\beta$ of the hw pseudo-SU(3) and of proxy-SU(3) are compared. 
In general, the predictions are very similar in the beginning of the neutron shell, 
while in the middle of the shell the proxy-SU(3) predictions are in general higher than the hw pseudo-SU(3) predictions. Finally, near the end of the shell, the hw pseudo-SU(3) predictions become higher than 
the proxy-SU(3) ones. In the same figure, predictions by Relativistic Mean Field (RMF) theory \cite{Lalazissis} are reported. In the beginning of the neutron shell the RMF predictions are lower 
than the hw pseudo-SU(3) and proxy-SU(3) predictions, while near midshell the RMF and proxy-SU(3) predictions are closer to each other. Near the end of the neutron shell the agreement between the three theories is better than in the beginning of the shell.  

 In Fig. (\ref{4G}) the predictions for $\gamma$ of the hw pseudo-SU(3) and of proxy-SU(3) are compared. 
Remarkable similarity between the results of the two theories is seen, despite the different approximations made and the different harmonic oscillator shells used in each of them. In particular, minima related to 
low values of $\mu$ in the $(\lambda,\mu)$ irrep appear for both theories around $N=100$-102 and $N=112$. 
Considerable disagreement is seen in the beginning of the neutron shell, where proxy-SU(3) predicts 
minima at $N=88$, 94, while the hw pseudo-SU(3) shows a stabilized region around $N=90$, related to the
gradual filling of the abnormal parity neutron orbitals. It should be remembered that the $N=90$ isotones $^{150}$Nd, $^{152}$Sm, and $^{154}$Gd are the best examples of the X(5) critical pointy symmetry \cite{IacX5}, characterizing the shape/phase transition \cite{McCutchan,Cejnar} between spherical and prolate deformed shapes.  

\section{Conclusion} 

In this work we have considered the implications of particle-hole symmetry breaking on the calculation of the collective shape variables $\beta$ and $\gamma$ within the framework of the pseudo-SU(3) scheme. The particle-hole symmetry breaking has deep physical roots, since it is due to the short range of the nucleon-nucleon interaction and the Pauli principle. Therefore it is expected to appear in a general way in systems of many fermions, thus paving the way for further investigations in many body fermionic systems in other branches of physics. A recent study in metallic clusters, explaining the appearance of prolate shapes above the magic numbers and of oblate shapes below the magic numbers seen in these physical systems will be published elsewhere \cite{PRA}.  Within the realm of the nuclear pseudo-SU(3) model, it has been shown that the particle-hole symmetry breaking leads to parameter independent predictions for the nuclear deformation $\beta$ which are in good agreement with relativistic and non-relativistic mean field predictions, as well as to the experimental data, where known. The parameter-independent predictions of the pseudo-SU(3) scheme for the $\gamma$ shape variable are even more dramatic. Particle-hole symmetry breaking leads to an answer to the long standing question of prolate over oblate deformation dominance in the ground states of even-even nuclei, as well as to the prediction of a prolate to oblate shape/phase transition in rare earths around 114-116 neutrons, in good agreement with available empirical information. The pseudo-SU(3) predictions are also in good agreement with the predictions of proxy-SU(3) \cite{proxy2,proxy3}. 
The compatibility of proxy-SU(3) and pseudo-SU(3) predictions has also been demonstrated recently in the study of quarteting in heavy nuclei \cite{Cseh}. It is interesting that these two different approximation methods of restoring the SU(3) symmetry in medium mass and heavy nuclei lead to results which are very similar to each other. Since pseudo-SU(3) has been applied mostly in the lower half of nuclear shells \cite{Vargas1,Vargas2,Popa}, the present work paves the way for its application in the upper half of nuclear shells, in which unique phenomena, as the prolate to oblate shape/phase transition take place.  

Restoration of an approximate SU(3) symmetry in nuclei beyond the sd shell can also be achieved in the framework of the quasi-SU(3) scheme \cite{Zuker1,Zuker2}, which has been found to be very appropriate for the description of $N=Z$ nuclei \cite{Zuker2}.  It would be an interesting project to examine how the particle-hole symmetry breaking appears within the quasi-SU(3) framework and how the prolate over oblate dominance and the prolate to oblate shape/phase transition come out within this model.  

\section*{Acknowledgements} 
Financial support by the Bulgarian National Science Fund (BNSF) under Contract No. KP-06-N28/6 is gratefully acknowledged. 

\section*{Author contribution statement}
 
All authors contributed equally to this article.


\begin{table*}[htb]

\caption{Distribution of valence protons and valence neutrons into normal and abnormal parity orbitals in the rare earth region, as obtained from the standard Nilsson diagrams \cite{Lederer}, using for each nucleus the deformation parameter obtained from Ref. \cite{Gogny}, as discussed in Section 2. In each sum,
the first number represents the normal parity nucleons, while the second number corresponds to the abnormal parity nucleons. See the two examples at the end of Section 2 for more detailed explanations. 
 The valence protons in normal parity orbitals live within a U(10) pseudo-SU(3) shell, while the valence neutrons in normal parity orbitals live within a U(15) pseudo-SU(3) shell. Adopted from Ref. \cite{Rila19}. 
}
\label{T1}

\bigskip

\rotatebox{90}{

\begin{tabular}{ r c c c c c c c c c c c c c  }

\hline\noalign{\smallskip}
  & Xe & Ba & Ce & Nd & Sm & Gd & Dy & Er & Yb & Hf & W & Os & Pt \\
Z$_{val}$ & 4  & 6  &  8 & 10 & 12 & 14 & 16 & 18 & 20 & 22 & 24 & 26 & 28 \\
   &4+0& 6+0 & 6+2&6+4 &8+4 &10+4&10+6&10+8&12+8&14+8&16+8&16+10&16+12\\

\noalign{\smallskip}\hline\noalign{\smallskip}
 N$_{val}$&     &     &     &     &     &     &     &     &     &     &     &     &     \\
 2&  2+0&  2+0&  2+0&  2+0&  2+0&  2+0&  2+0&  2+0&  2+0&  2+0&  2+0&  2+0&  2+0\\
 4&  4+0&  4+0&  4+0&  4+0&  4+0&  4+0&  4+0&  4+0&  4+0&  4+0&  4+0&  4+0&  4+0\\
 6&  6+0&  6+0&  6+0&  6+0&  6+0&  6+0&  6+0&  6+0&  6+0&  6+0&  6+0&  6+0&  6+0\\
 8&  8+0&  6+2&  6+2&  6+2&  6+2&  6+2&  6+2&  6+2&  6+2&  8+0&  8+0&  8+0&  8+0\\
10&  8+2&  6+4&  6+4&  6+4&  6+4&  6+4&  6+4&  6+4&  6+4&  8+2&  8+2&  8+2& 10+0\\
12&  8+4&  8+4&  8+4&  8+4&  8+4&  8+4&  8+4&  8+4&  8+4& 10+2&  8+4&  8+4& 10+2\\
14& 10+4&  8+6&  8+6&  8+6&  8+6&  8+6&  8+6&  8+6&  8+6& 10+4& 10+4&  8+6& 10+4\\
16& 10+6& 10+6& 10+6& 10+6& 10+6& 10+6& 10+6& 10+6& 10+6& 12+4& 10+6& 10+6& 10+6\\
18& 12+6& 12+6& 12+6& 12+6& 12+6& 12+6& 12+6& 12+6& 12+6& 12+6& 12+6& 12+6& 12+6\\
20& 12+8& 12+8& 12+8& 12+8& 12+8& 12+8& 12+8& 12+8& 14+6& 14+6& 12+8& 12+8& 12+8\\
22& 14+8& 14+8& 14+8& 14+8& 14+8& 14+8& 14+8& 14+8& 14+8& 16+6& 14+8& 14+8& 14+8\\
24& 16+8& 16+8& 16+8& 16+8& 16+8& 16+8& 16+8& 16+8& 16+8& 16+8& 16+8&14+10& 16+8\\
26&16+10&16+10&16+10&16+10&16+10&16+10&16+10&16+10& 18+8&16+10&16+10&16+10&16+10\\
28&18+10&18+10&18+10&18+10&18+10&18+10&18+10&18+10&18+10&18+10&18+10&18+10&18+10\\
30&20+10&20+10&20+10&20+10&20+10&20+10&20+10&20+10&20+10&20+10&20+10&20+10&20+10\\
32&22+10&22+10&22+10&22+10&22+10&22+10&22+10&22+10&22+10&22+10&22+10&22+10&22+10\\
34&24+10&24+10&24+10&24+10&24+10&24+10&24+10&24+10&24+10&24+10&24+10&24+10&24+10\\
36&24+12&26+10&24+12&24+12&24+12&24+12&24+12&24+12&24+12&24+12&24+12&26+10&24+12\\
38&24+14&26+12&24+14&24+14&24+14&24+14&24+14&24+14&24+14&24+14&24+14&26+12&24+14\\
40&26+14&26+14&26+14&26+14&26+14&26+14&26+14&26+14&26+14&26+14&26+14&26+14&26+14\\
42&28+14&28+14&28+14&28+14&28+14&28+14&28+14&28+14&28+14&28+14&28+14&28+14&28+14\\

\noalign{\smallskip}\hline
                                                                    
\end{tabular}

}

\end{table*} 

\begin{table*}[htb]

\caption{Highest weight irreducible representations (irreps), labeled by hw, and irreps possessing the highest eigenvalue of the second order Casimir operator of SU(3) (see Eq. (\ref{C2})), labeled by hC,   
occurring in the decomposition of U(10) and U(15) for $M$ particles, as obtained through the code of Ref. \cite{code}. Oblate irreps are shown in boldface. A more extended version of the table has been given in Ref. \cite{proxy2}. Adopted from Ref. \cite{Rila19}. 
}
\label{T2}

\bigskip
\begin{tabular}{ r c c c c c c c   }

\hline\noalign{\smallskip}
  M     & 2 & 4 & 6 & 8 & 10 & 12 & 14 \\
\noalign{\smallskip}\hline\noalign{\smallskip}
U(10) hw&(6,0)&(8,2)&(12,0)&(10,4)&(10,4)&(12,0)&(6,6)\\
U(10) hC &(6,0)&(8,2)&(12,0)&(10,4)&(10,4)&{\bf (4,10)}&{\bf(0,12)}\\
U(15) hw&(8,0)&(12,2)&(18,0)&(18,4)&(20,4)&(24,0)&(20,6)\\
U(15) hC &(8,0)&(12,2)&(18,0)&(18,4)&(20,4)&(24,0)&(20,6)\\
\noalign{\smallskip}\hline

\end{tabular}

\begin{tabular}{ r c c c c c c c   }

\hline\noalign{\smallskip}
  M     &   16 & 18 & 20 & 22 & 24 & 26 & 28  \\
\noalign{\smallskip}\hline\noalign{\smallskip}
U(10) hw & {\bf (2,8)}&{\bf(0,6)}&(0,0)&   &   &  &   \\
U(10) hC & {\bf(2,8)}&{\bf(0,6)}&(0,0)&  &   &  &   \\
U(15) hw & (18,8)&(18,6)&(20,0)&(12,8)&{\bf (6,12)}&{\bf(2,12)}&{\bf(0,8)}\\
U(15) hC &{\bf (6,20)}&{\bf(0,24)}&{\bf(4,20)}&{\bf(4,18)}&{\bf(0,18)}&{\bf(2,12)}&{\bf(0,8)}\\
\noalign{\smallskip}\hline

\end{tabular}

\end{table*}

\begin{table*}[htb]

\caption{Total irreps corresponding to rare earth nuclei obtained when the highest weight irreps (hw) are used for both the valence protons and the valence neutrons. 
The irreps are taken from Table 2, as explained in Section 2 through two examples. Oblate irreps are shown in boldface. Adopted from Ref. \cite{Rila19}. 
}
\label{T3}

\bigskip

\rotatebox{90}{

\begin{tabular}{ r c c c c c c c c c c c c c }

\hline\noalign{\smallskip}
N & Xe & Ba & Ce & Nd & Sm & Gd & Dy & Er & Yb & Hf & W & Os & Pt \\
\noalign{\smallskip}\hline\noalign{\smallskip}
 84& 16,2& 20,0& 20,0& 20,0& 18,4& 18,4& 18,4& 18,4& 20,0& 14,6& 10,8& 10,8& 10,8\\
 86& 20,4& 24,2& 24,2& 24,2& 22,6& 22,6& 22,6& 22,6& 24,2& 18,8&14,10&14,10&14,10\\
 88& 26,2& 30,0& 30,0& 30,0& 28,4& 28,4& 28,4& 28,4& 30,0& 24,6& 20,8& 20,8& 20,8\\
 90& 26,6& 30,0& 30,0& 30,0& 28,4& 28,4& 28,4& 28,4& 30,0&24,10&20,12&20,12&20,12\\
 92& 26,6& 30,0& 30,0& 30,0& 28,4& 28,4& 28,4& 28,4& 30,0&24,10&20,12&20,12&22,12\\
 94& 26,6& 30,4& 30,4& 30,4& 28,8& 28,8& 28,8& 28,8& 30,4&26,10&20,12&20,12&22,12\\
 96& 28,6& 30,4& 30,4& 30,4& 28,8& 28,8& 28,8& 28,8& 30,4&26,10&22,12&20,12&22,12\\
 98& 28,6& 32,4& 32,4& 32,4& 30,8& 30,8& 30,8& 30,8& 32,4& 30,6&22,12&22,12&22,12\\
100& 32,2& 36,0& 36,0& 36,0& 34,4& 34,4& 34,4& 34,4& 36,0& 30,6& 26,8& 26,8& 26,8\\
102& 32,2& 36,0& 36,0& 36,0& 34,4& 34,4& 34,4& 34,4& 32,6&26,12& 26,8& 26,8& 26,8\\
104& 28,8& 32,6& 32,6& 32,6&30,10&30,10&30,10&30,10& 32,6&24,14&22,14&22,14&22,14\\
106&26,10& 30,8& 30,8& 30,8&28,12&28,12&28,12&28,12& 30,8&24,14&20,16&22,14&20,16\\
108&26,10& 30,8& 30,8& 30,8&28,12&28,12&28,12&28,12& 30,6&24,14&20,16&20,16&20,16\\
110& 26,8& 30,6& 30,6& 30,6&28,10&28,10&28,10&28,10& 30,6&24,12&20,14&20,14&20,14\\
112& 28,2& 32,0& 32,0& 32,0& 30,4& 30,4& 30,4& 30,4& 32,0& 26,6& 22,8& 22,8& 22,8\\
114&20,10& 24,8& 24,8& 24,8&22,12&22,12&22,12&22,12& 24,8&18,14&{\bf 14,16}&{\bf 14,16}&{\bf 14,16}\\
116&14,14&18,12&18,12&18,12&16,16&16,16&16,16&16,16&18,12&{\bf 12,18}&{\bf 8,20}&{\bf 8,20}&{\bf 8,20}\\
118&14,14&14,12&18,12&18,12&16,16&16,16&16,16&16,16&18,12&{\bf 12,18}&{\bf 8,20}&{\bf 4,20}&{\bf 8,20}\\
120&14,14&14,12&18,12&18,12&16,16&16,16&16,16&16,16&18,12&{\bf 12,18}& {\bf 8,20}&{\bf  4,20}&{\bf 8,20}\\
122&{\bf 10,14}&14,12&14,12&14,12&{\bf 12,16}&{\bf 12,16}&{\bf 12,16}&{\bf 12,16}&14,12& {\bf 8,18}& {\bf 4,20}&{\bf 4,20}&{\bf 4,20}\\
124&{\bf 8,10}& 12,8& 12,8& 12,8&{\bf 10,12}&{\bf 10,12}&{\bf 10,12}& {\bf 10,12}& 12,8& {\bf 6,14}&{\bf 2,16}& {\bf 2,16}& {\bf 2,16}\\

\noalign{\smallskip}\hline

\end{tabular}

}

\end{table*} 

\begin{table*}[htb]

\caption{Total irreps corresponding to rare earth nuclei obtained when the irrep having the highest eigenvalue  of the second order Casimir operator of SU(3) (hC)
is used for both the valence protons and the valence neutrons. The irreps are taken from Table 2, as explained in Section 2 through two examples. Oblate irreps are shown in boldface. Adopted from Ref. \cite{Rila19}. 
}
\label{T4}

\bigskip

\rotatebox{90}{

\begin{tabular}{ r c c c c c c c c c c c c c }

\hline\noalign{\smallskip}
N & Xe & Ba & Ce & Nd & Sm & Gd & Dy & Er & Yb & Hf & W & Os & Pt \\
\noalign{\smallskip}\hline\noalign{\smallskip}
 
 84& 16,2& 20,0& 20,0& 20,0& 18,4& 18,4& 18,4& 18,4&12,10& 8,12& 10,8& 10,8& 10,8\\
 86& 20,4& 24,2& 24,2& 24,2& 22,6& 22,6& 22,6& 22,6&16,12&12,14&14,10&14,10&14,10\\
 88& 26,2& 30,0& 30,0& 30,0& 28,4& 28,4& 28,4& 28,4&22,10&18,12& 20,8& 20,8& 20,8\\
 90& 26,6& 30,0& 30,0& 30,0& 28,4& 28,4& 28,4& 28,4&22,10&18,16&20,12&20,12&20,12\\
 92& 26,6& 30,0& 30,0& 30,0& 28,4& 28,4& 28,4& 28,4&22,10&18,16&20,12&20,12&22,12\\
 94& 26,6& 30,4& 30,4& 30,4& 28,8& 28,8& 28,8& 28,8&22,14&20,16&20,12&20,12&22,12\\
 96& 28,6& 30,4& 30,4& 30,4& 28,8& 28,8& 28,8& 28,8&22,14&20,16&22,12&20,12&22,12\\
 98& 28,6& 32,4& 32,4& 32,4& 30,8& 30,8& 30,8& 30,8&24,14&24,12&22,12&22,12&22,12\\
100& 32,2& 36,0& 36,0& 36,0& 34,4& 34,4& 34,4& 34,4&28,10&24,12& 26,8& 26,8& 26,8\\
102& 32,2& 36,0& 36,0& 36,0& 34,4& 34,4& 34,4& 34,4&24,16&20,18& 26,8& 26,8& 26,8\\
104& 28,8& 32,6& 32,6& 32,6&30,10&30,10&30,10&30,10&24,16&{\bf 6,32}&22,14&22,14&22,14\\ 
106&{\bf 14,22}&{\bf 18,20}&{\bf 18,20}&{\bf 18,20}&{\bf 16,24}&{\bf 16,24}&{\bf 16,24}&{\bf 16,24}&{\bf 10,30}& {\bf 6,32}& {\bf 8,28} &{\bf 22,14}& {\bf 8,28}\\
108&{\bf 14,22} &{\bf 18,20} &{\bf 18,20}&{\bf 18,20}&{\bf 16,24}&{\bf 16,24}&{\bf 16,24}&{\bf 16,24}&{\bf 4,34}&{\bf 6,32}&{\bf 8,28}&{\bf 8,28}&{\bf 8,28}\\
110&{\bf 8,26}&{\bf 12,24}&{\bf 12,24}&{\bf 12,24}&{\bf 10,28}&{\bf 10,28}&{\bf 10,28}&{\bf 10,28}&{\bf  4,34}&{\bf 0,36}&{\bf 2,32}&{\bf 2,32}&{\bf 2,32}\\
112&{\bf 12,22}&{\bf 16,20}&{\bf 16,20}&{\bf 16,20}&{\bf 14,24}&{\bf 14,24}&{\bf 14,24}&{\bf 14,24}&{\bf  8,30}&{\bf 4,32}&{\bf 6,28}&{\bf 6,28}&{\bf  6,28}\\
114&{\bf 12,20}&{\bf 16,18}&{\bf 16,18}&{\bf 16,18}&{\bf 14,22}&{\bf 14,22}&{\bf 14,22}&{\bf 14,22}&{\bf  8,28}& {\bf 4,30}& {\bf 6,26}& {\bf 6,26}&{\bf 6,26}\\
116&{\bf 8,20}&{\bf 12,18}&{\bf 12,18}&{\bf 12,18}&{\bf 10,22}&{\bf 10,22}&{\bf 10,22}&{\bf 10,22}&{\bf 4,28}&{\bf 0,30}&{\bf 2,26}&{\bf 2,26}&{\bf 2,26}\\
118&{\bf 8,20}&14,12&{\bf 12,18}&{\bf 12,18}&{\bf 10,22}&{\bf 10,22}&{\bf 10,22}&{\bf 10,22}&{\bf 4,28}& {\bf 0,30}&{\bf 2,26}&{\bf 4,20}&{\bf 2,26}\\
120&{\bf 8,20}&14,12&{\bf 12,18}&{\bf 12,18}&{\bf 10,22}&{\bf 10,22}&{\bf 10,22}&{\bf 10,22}&{\bf 4,28}& {\bf 0,30}&{\bf 2,26}&{\bf 4,20}&{\bf 2,26}\\
122&{\bf 10,14}&14,12&14,12&14,12&{\bf 12,16}&{\bf 12,16}&{\bf 12,16}&{\bf 12,16}&{\bf 6,22}&{\bf 2,24}& {\bf 4,20}&{\bf 4,20}&{\bf 4,20}\\
124&{\bf 8,10}& 12,8& 12,8& 12,8&{\bf 10,12}&{\bf 10,12}&{\bf 10,12}&{\bf 10,12}&{\bf 4,18}&{\bf 0,20}& {\bf 2,16}&{\bf 2,16}&{\bf 2,16}\\

\noalign{\smallskip}\hline

\end{tabular}

}

\end{table*}


\end{document}